\documentclass[12pt,preprint]{aastex}

\usepackage{pdflscape}

\slugcomment{To appear in ApJ}

\shorttitle{Dust around RCB stars. II}
\shortauthors{Garc\'{\i}a-Hern\'andez et al.}

\begin{document}


\title{Dust around R Coronae Borealis stars: II. Infrared emission features in a H-poor
environment}


\author{D. A. Garc\'{\i}a-Hern\'andez\altaffilmark{1,2}, N. Kameswara
Rao\altaffilmark{3,4}, D. L. Lambert\altaffilmark{4}}


\altaffiltext{1}{Instituto de Astrof\'{\i}sica de Canarias, C/ Via L\'actea s/n, 38200 La Laguna, Spain; agarcia@iac.es}
\altaffiltext{2}{Departamento de Astrof\'{\i}sica, Universidad de La Laguna (ULL), E-38206 La Laguna, Tenerife, Spain}
\altaffiltext{3}{543, 17$^{th}$ Main, IV Sector, HSR Layout, Bangalore 560102 and Indian Institute of Astrophysics, Bangalore 560034, India; nkrao@iiap.res.in}
\altaffiltext{4}{W. J. McDonald Observatory. The University of Texas at Austin. 1 University Station, C1400. Austin, TX 78712$-$0259, USA; dll@astro.as.utexas.edu}


\begin{abstract}
Residual {\it Spitzer/IRS} spectra for a sample of 31 R Coronae Borealis
(RCB) stars are presented and discussed in terms of narrow emission features
superimposed on the quasi-blackbody continuous infrared emission. A broad
$\sim$6-10 $\mu$m dust emission complex is seen in the RCBs showing an extreme
H-deficiency. A secondary and much weaker $\sim$11.5-15 $\mu$m broad emission
feature is detected in a few RCBs with the strongest $\sim$6-10 $\mu$m dust
complex.  The {\it Spitzer} infrared spectra reveal for the first time the
structure within the $\sim$6-10 $\mu$m dust complex, showing the presence of
strong C-C stretching modes at $\sim$6.3 and 8.1 $\mu$m as well as of other dust
features at $\sim$5.9, 6.9, and 7.3 $\mu$m, which are attributable to amorphous
carbonaceous solids with little or no hydrogen. The few RCBs with only
moderate H-deficiencies display the classical `unidentified infrared bands
(UIRs)' and mid-infrared features from fullerene-related molecules. In general,
the characteristics of the RCB infrared emission features are not correlated
with the stellar and circumstellar properties, suggesting that the RCB dust
features may not be dependent on the present physical conditions around RCB
stars. The only exception seems to be the central wavelength of the 6.3 $\mu$m
feature, which is blue-shifted in those RCBs showing also the UIRs, i.e., the
RCBs with the smallest H deficiency.
\end{abstract}





\section{Introduction}

The R Coronae Borealis (hereafter RCB) stars are  peculiar stars with two main
distinguishing characteristics:  an extreme hydrogen deficiency (ranging from
factors of $\sim$10-100 to greater than 10$^8$)  and an unpredictable optical
variability with relatively rapid declines in brightness (variations from 2 to 8
magnitudes are usually seen in the V-band) that can last from a few weeks to
many months (see e.g., Lambert \& Rao 1994; Clayton 1996, 2012). The
H-deficiency seen in RCBs may be explained by two possible scenarios: the
double-degenerate (DD) and the final flash (FF) scenarios. The resulting
H-deficient star is produced either  from the merger of a He  white dwarf with a
C-O white dwarf (DD scenario)\footnote{Note that Zhang \& Jeffery (2012) suggest
that some RCB stars may be formed from the merger of two He white dwarfs.}
or from a final post-asymptotic giant branch (post-AGB) helium flash in the
central stars of Planetary Nebulae (FF scenario). Detailed abundance analysis of
RCBs (Lambert \& Rao 1994; Asplund et al. 2000; Clayton et al. 2005, 2007;
Garc\'{\i}a-Hern\'andez et al. 2009, 2010a; Jeffery et al. 2011; Pandey \&
Lambert 2011) suggest that the DD scenario may account for the great majority of
the RCBs. Indeed, simulations by Longland et al. (2011) show that  a ``hot"
white dwarf merger may qualitatively explain the chemical abundances observed in
RCB stars\footnote{In this context, `hot' means that nucleosynthesis occurs
during and following a merger in contrast to a `cold' merger where material is
mixed but unaltered by nucleosynthesis.} (see also Staff et al. 2012 for
more details about merger simulations and RCBs). Another simulation of a hot
merger and its subsequent post-merger evolution, nucleosynthesis and mixing
(Menon et al. 2012), provides a rather satisfactory account of observed
abundances for RCBs. 

A RCB's characteristic optical variability is caused by the formation of dust
clouds along the line of sight towards the star (e.g., Forrest, Gillett \& Stein
1972; Le\~ao et al. 2007). A collection of dust clouds around the RCB through
absorption and reradiation of stellar flux  provide an infrared excess. A
typical RCB star emits approximately 30\% of the stellar flux in the infrared,
confirming that they are producers of dust with typical blackbody temperatures
ranging from $\sim$400 K to 900 K (Stein et al. 1969; Feast et al.
1997)\footnote{Note that a colder ($\sim$30-100 K) dust component, has also been
detected in some RCB stars (e.g., Rao \& Nandy 1986). Cold dust around R
CrB itself has been recently detected out to 500 $\mu$m (Clayton et al.
2011a).}. However, key questions about the composition of the dust grains around
RCBs as well as about where and how (i.e., the possible chemical pathways
followed in the condensation process) dust nucleation takes place remain
unanswered (see also Garc\'{\i}a-Hern\'andez, Rao \& Lambert 2011a). This has
been due to the lack of high-quality infrared spectra for all but the two or
three infrared-brightest RCB stars. 

Recently, we have tried to correct this situation by carrying out a {\it
Spitzer/IRS} spectroscopic survey for essentially all of  the known Galactic 
RCB stars (Garc\'{\i}a-Hern\'andez, Rao \& Lambert 2011a, hereafter Paper I). 
Examination of RCB's infrared spectra and detection of spectroscopic features is
a pathway to  identification of the circumstellar constituents (dust grains and
large molecules). {\it Spitzer/IRS} spectra of RCBs are especially valuable
because, in contrast to the usual dusty red giants (and main sequence stars),
dust in a RCB is formed in a H-poor, He-rich, and usually C-rich environment: a
C/He ratio of one per cent by number and a C/O ratio of somewhat  greater than
one are typical (Asplund et al. 2000). An understanding of dust formation in
both  H-rich (both O-rich and C-rich) normal stars and H-poor (C-rich)
environments of the RCBs is desired before one can pronounce that dust formation
is understood; see Woitke et al. (1996) for a quantitative model of carbon
chemistry and dust condensation around RCB stars.

In Paper I, individual spectral energy distributions (SEDs) compiled from
optical and infrared photometry and the {\it Spitzer} spectra were corrected for
interstellar reddening and then fit by blackbodies representing the star and the
dust. Subtraction of the flux provided by the blackbodies from  the dereddened
SED provides what we term the residual or difference spectrum which is dominated
by infrared emission features. In this second paper of the series of {\it
Spitzer/IRS} RCB spectra, we discuss the residual (or difference) spectra. The
residual {\it Spitzer/IRS} spectra of the least H-poor RCBs DY Cen and V854 Cen
showing both possible C$_{60}$ or fullerene-related features and the classical
`unidentified infrared bands (UIRs)' - usually attributed to polycyclic aromatic
hydrocarbons (PAHs) - have been previously discussed by us
(Garc\'{\i}a-Hern\'andez, Rao \& Lambert 2011b) but see Garc\'{\i}a-Hern\'andez,
Rao \& Lambert (2012). Clayton et al. (2011b)  recently discussed residual {\it
Spitzer} spectra of the two hot RCB stars V348 Sgr and HV 2671 and noted that
the spectrum of the former is very different from that of the latter.

A summary of the {\it Spitzer/IRS} observations and of the construction of the
residual spectra is presented in Section 2. Section 3 gives an
overview of the {\it Spitzer/IRS} residual spectra, where the dust
features are easily seen, while Section 4 discusses the characteristics and
identifications of the dust features around RCB stars.  A few RCBs with
exceptional residual spectra are highlighted in Section 5. The
conclusions of our work are given Section 6. In a future paper we will attempt
to model the infrared RCB spectra in order  to get information on the physical
properties of the dust clouds.

\section{{\it Spitzer/IRS} observations and residual spectra}

{\it Spitzer Space Telescope} observations of a large number of RCB stars were
presented and discussed in Paper I. Stars  were chosen to provide comprehensive
coverage of warm RCBs across the chemical composition range observed in RCB
stars, to sample the coolest RCBs as well as to complete observations of
minority RCBs.\footnote{Minority RCBs were defined by Lambert \& Rao (1994) as
stars having extraordinarily high Si/Fe and S/Fe  ratios.}  Table 1 (updated
from Paper I) lists the 31 RCB stars included in our study together with some
relevant information such as stellar and dust temperatures, the interstellar
reddening E(B-V), and variability status. Finally, the inner parts of the
hydrogen deficient planetary nebulae (PNe) A 78, A 30, and IRAS 1833-2357 have
also been observed by {\it Spitzer} and we retrieved their infrared spectra from
the {\it Spitzer} database for comparison purposes.

The stellar effective temperature ($T_{\star}$) is taken from Paper I. The
blackbody temperatures of the dust continua are given as $T_{BB1}$ and, where
necessary, as $T_{BB2}$ with the corresponding fractional coverages as $R_{BB1}$
and $R_{BB2}$\footnote{The fractional covering factors $R_{BB1,2}$ are
defined as the flux ratios for the blackbodies relative to the stellar flux
$R_{BB1,2}$ = $f_{BB1,2}$/$f_{star}$ (see Paper I for more details).}. The
equivalent width of the 6-10 $\mu$m emission feature is given in the column
headed EQW$_{6-10\mu m}$ (see below for remarks on how this quantity was
measured). The final two columns describe the optical brightness of the star at
the time of the {\it Spitzer} observations and give the estimated interstellar
reddening from Paper I.

We refer the reader to Paper I for a detailed description of the  SEDs from
$\sim$0.4 to 40 $\mu$m for the  RCB stars in our sample  and the construction of
the blackbody fits to the continuum emitted by the circumstellar 
dust.\footnote{Note that a star's photospheric contribution to the observed {\it
Spitzer} spectra is generally very small.}  Subtraction of the blackbody fluxes 
from the reddening-corrected infrared fluxes provides the residual spectrum for
a RCB.  Blackbody continua are essentially a convenient device for specifying
the local continuum on which sit the emission features. Interpretation of the
spectral structure in the residual spectra is the principal focus of this
paper. 

The shape of this spectral structure is sensitive to several factors. First, the
correction for interstellar reddening  and especially the correction for the
interstellar 9.7 $\mu$m (and 18 $\mu$m) amorphous silicate features affects the
long-wavelength shape of the 6-10 $\mu$m dominant feature in the residual
spectra. Detailed information on the effect of reddening corrections is given in
the next Section. The method of correction assumes that the  strength of the
amorphous silicate features relative to the broad wavelength-dependent
extinction is the same for all lines of sight and independent of E(B-V). Second,
the shape of the short-wavelength end of the 6-10 $\mu$m dominant feature is
affected by the gap in contemporary measurements defining the SED between the
short-wavelength limit of the {\it Spitzer} observations and the K-band. Flux
from the  fitted dust blackbody is falling rapidly shortward of about 6 $\mu$m
but a contribution across the 3-6 $\mu$m interval from warmer dust  closer to
the star is ill-defined for lack of  photometry and particularly
spectrophotometry across the L-M-N bands at the time of the {\it Spitzer}
observations.  Thirdly, the discernible detail across the residual spectra is
dependent - of course - on the signal-to-noise (S/N) ratio of the {\it Spitzer}
spectra. 

In Paper I, the interstellar reddening was estimated in a very inhomogeneous
way with emphasis given to estimates from various techniques reported in
previous papers discussing the individual stars. Thus, it is of great interest
to compare these estimates with independent estimates obtained in a very
homogeneous way. Such estimates have been provided by Tisserand (2012) from the
COBE/DIRBE maps (Schlegel et al. 1998). In Figure 1, we compare Tisserand's
E(B-V) with those from Paper I. The dashed line shows that the two methods give
generally very similar values, especially for E(B-V) less than about 0.5. Two
obvious discrepancies at large E(B-V) -- FH Sct and UV Cas -- are marked on the
figure (see Section 2.1 for more details). 

In order to reduce the influence of the reddening correction (Section 2.1) on
the residual spectra, we consider first and foremost those RCBs with the
smallest E(B-V) values, say E(B-V) $\leq 0.30$ where E(B-V) is taken from Table
1. The octet includes S Aps, Z UMi, WX CrA, V1157 Sgr, VZ Sgr, U Aqr, V CrA, and
RS Tel.  R CrB and RY Sgr would be in this group but for the fact that their
{\it Spitzer} spectra do not extend shortward of 10 $\mu$m; we are
reluctant to combine {\it Infrared Space Observatory} (ISO) spectra (Lambert et
al. 2001) with the {\it Spitzer} spectra for these two stars which are so
obviously variable in the mid-infrared.  V854 Cen and HV 2671, also low E(B-V)
stars, and DY Cen, a star with a higher E(B-V),  have distinctly different
residual spectra, presumably on account of their small H-deficiencies. Their
residual spectra were discussed by Garc\'{\i}a-Hern\'{a}ndez, Rao \& Lambert
(2011b). Fortunately, the low-reddening octet samples the different classes
(warm vs cool, majority vs minority) of RCBs.  Thus, it is likely that their
residual spectra are representative  of  the range among RCBs (see below). 
Construction of residual spectra is illustrated in Figure 2 for S Aps (E(B-V) =
0.05) and V1157 Sgr (E(B-V) = 0.30). Figure 3 shows the complete set of residual
spectra for the low reddening octet. The residual spectra are normalized to the
maximum value longward of 6 $\mu$m (i.e., the occasional and spurious flux rise
at the short wavelength limit seen in some stars is ignored) and displaced for
clarity.  Residual spectra for the more-reddened RCBs are shown in Figures 4, 5,
and 6. V517 Oph and SU Tau are not shown in these Figures because their {\it
Spitzer} spectra do not extend shortward of 10 $\mu$m (see Section 3). The
effect on the residual spectra of reddening uncertainties is discussed below.

An artefact in the {\it Spitzer} spectra should be noted. Some RCB stars (e.g.,
V348 Sgr, RZ Nor, and V CrA)  display an emission feature centred at
$\sim$14.2$\pm$0.2 $\mu$m.  Although the feature may be attributable to a weak
C-C-C bending mode (Hony et al. 2001) in carbon grains, we noted that detection
of this feature is made using the low-resolution modules, and the feature lies
right on the boundary between the SL and LL modules. In addition, this feature
is not obvious in the SH spectra of the few sources (e.g., V854 Cen) observed 
with both low- and high-resolution modules. According to the IRS Instrument
Handbook\footnote{see e.g.,
http://ssc.spitzer.caltech.edu/irs/irsinstrumenthandbook/home/}, extreme caution
is recommended in interpreting features at 13.2-15 $\mu$m because there is a
known ``SL 14 micron teardrop", which produces an excess emission that is
correlated with the source brightness. The apparent 14.2 $\mu$m feature is
stronger in the brightest RCBs, and, thus, we  conclude that the 14.2 $\mu$m
feature is primarily an artifact related to the ``SL 14 micron teardrop".

\subsection{Correction for interstellar reddening}

We have undertaken an examination of the effect of the  correction for
interstellar reddening (i.e., varying E(B-V)) on the resulting spectral
structure (e.g., the shape of the 6-10 $\mu$m emission complex). As we mentioned
above, our estimates of E(B-V) values from Paper I are compared with the
independent estimates by Tisserand (2012) as obtained from COBE/DIRBE
measurements, for 29 of the 31 stars in Table 1; Tisserand did not give E(B-V)
for MACHOJ181933 and HV 2671. For these 29 RCB stars, the mean difference 
between Tisserand's E(B-V) and our E(B-V) values is 0.044 with a root mean
square deviation of 0.14.  We thus assume that the error in our E(B-V) estimates
is about 0.1 (see Figure 1). 

We have adopted the reddening curve of Chiar \& Tielens (2006) of the diffuse
interstellar medium (ISM) (their Table 7). This table provides the extinction 
from 1.24 $\mu$m to 30 $\mu$m but was extended to 38 $\mu$m, as indicated in
Paper I. The 9.7 $\mu$m (and 18 $\mu$m) profile of the silicate feature inferred
from observations of the Wolf-Rayet (WR) star WR98a is taken as representative
of the local ISM.  Our adoption of a single profile for the 9.7 and 18 $\mu$m
silicate features might affect the reddening-corrected emission profiles in the
$\sim$8-11 $\mu$m range  because the silicate profile (as well as other profile
changes at 14 $\mu$m) varies towards dark clouds, the Galactic center and/or
other directions. The strength of the 9.7 $\mu$m  feature also changes
appreciably even in the diffuse ISM. This is reflected in the
A$_{v}$/$\tau$(9.7) ratio, which changes from 14.4 to 17.5 even for the
calibrating WR stars used by Chiar \& Tielens (2006) and Williams et al. (2012)
give A$_{v}$/$\tau$(9.7)=16.6$\pm$4.6 for WR stars. 

Correction for the silicate feature centered at 9.7 $\mu$m  affects the 6-10
$\mu$m profile from about 8.2 to 11.2 $\mu$m. Obviously, increasing the
reddening increases the emission for $\sim$8 to 11 $\mu$m. This is clearly shown
in Figures 7 and 8 where we display resultant residual spectra obtained for
various E(B-V) values for V3795 Sgr and S Aps.  In the case of V3795 Sgr (Figure
7), our adopted E(B-V) of 0.79 is preferred because the optical colors are
better corrected with this E(B-V) value. E(B-V) values in the range
$\sim$0.65-0.82 may fit reasonably well the {\it Spitzer} spectrum, being
consistent with our estimated E(B-V) uncertainty of about 0.1. Unlike the
majority of RCBs, the emission excess for V3795 Sgr relative to the fitted
blackbody extends to about 15 $\mu$m but the profile of this extension is
effectively independent of the assumed E(B-V) value. Figure 8 shows the
sensitivity of the residual spectra for S Aps, a star almost unreddened by the
intervening interstellar medium: spectra are shown for E(B-V) of 0.0 (as
observed), 0.2, and 0.4. For both stars V3795 Sgr and S Aps, the dust
temperature obtained from blackbody fits to the {\it Spitzer} continua are
essentially unaffected by the adopted E(B-V) values over the range investigated.
The total flux emitted in the infrared is weakly dependent on E(B-V). The
principal effect of the reddening is its effect on the profile of the (dust)
6-12 $\mu$m emission feature where correction for the interstellar silicate
feature defines the profile  between about 8.2 and 11.2 $\mu$m.    

\subsection{Characterization of emission components}

Characterization of the emission features in the residual spectra begins with
estimating the equivalent width (EQW) of the $\sim$6-10 $\mu$m dust emission
complex seen in all RCB stars in our sample (see Figures 3$-$6). For this
measurement, the wavelength interval was adjusted for the individual star. A
source of uncertainty arises from the definition of the short-wavelength edge to
the broad 6-10$\mu$m emission feature (see above). The EQWs (Table 1 and the
column headed EQW$_{6-10\mu m}$) appear to fall into two groups. In the minor
group with large EQW are the less H-deficient RCBs DY Cen, V854 Cen, and HV
2671\footnote{We note that the chemical abundances (e.g., H-content) of HV
2671 are not known but its optical spectrum is very similar to V348 Sgr (De
Marco et al. 2002). The abundances of both stars are thought to resemble those
from [WC] central stars but their {\it Spitzer} spectra are very different
(Clayton et al. 2011b). The HV 2671 {\it Spitzer} spectrum (with PAH-like
features) looks identical to the other less H-deficient RCBs, while the V348 Sgr
IR spectrum is more similar to that of extremely H-poor RCBs. Thus, based on the
{\it Spitzer} spectra alone, we consider HV 2671 as a possibly less H-deficient
star.} and also the minority star V3795 Sgr and the hot RCB MV Sgr. The major
group with EQWs less than about 0.6 $\mu$m are the `normal' RCBs. This apparent
separation and the EQW spread within the major group are discussed in the next
section.

Residual {\it Spitzer} spectra display structure within the 6-10
$\mu$m dust emission complex of RCB stars. In spectra of high S/N ratio (e.g.,
UW Cen), the complex is seemingly resolvable into emission components at
$\sim$5.9, 6.3, 6.9, 7.3, 7.7, 8.1, 8.6, 9.1, and 9.6 $\mu$m. These emission
components have similar wavelengths and similar widths (FWHM) across the sample
(see below). In most RCBs, however, the features are blended and discerning them
may not be an easy task, particularly between 7 and 10 $\mu$m (i.e., other than
the 6.3 $\mu$m feature). 
%It is to be noted here that most RCB spectra,
%displaying the strongest features at 6.3 $\mu$m and at about 8 $\mu$m, would
%fall in `class C' of the observed astrophysical UIR spectra as defined by
%Peeters et al. (2002). 
High S/N residual RCB spectra
such as UW Cen set the central wavelength of the emission components within the
broad 7-10 $\mu$m feature and show that there are several components rather than
one broad asymmetrical band. Indeed, by overlapping most of the
residual RCB spectra on each other, common emission bumps at
$\sim$6.3, 6.9, 7.3, 7.7, 8.1, and 8.6 $\mu$m are usually found. Thus, we were
encouraged to carry out a multi-Gaussian fit using a deblending routine 
written for
SUPERMONGO. A condition of the fit was that Gaussians were introduced at
wavelengths of 5.9, 6.3, 7.3, 7.8, 8.1, 8.6, 9.1, and 9.6 $\mu$m with small
adjustments allowed for the central wavelengths and FWHMs. Parameters from the
multi-Gaussian fits (central wavelength, FWHM, and integrated flux) are listed
in Table 2 for the low-reddening octet and in Table 3 for the more reddened
stars including the three least H-deficient RCBs. Examples of fits are shown in
Figure 9 for UW Cen and V348 Sgr, a warm and a hot RCB, respectively. In
general, the FWHM of components in the 7-14 $\mu$m interval are considerably
greater than the instrumental width which according to the {\it Spitzer} manual
is two pixels or 0.12
$\mu$m\footnote{http://irsa.ipac.caltech.edu/data/SPITZER/docs/irs/irsinstrumenthandbook/}     
The 5.9 $\mu$m component has a FWHM $\sim$ 0.2 $\mu$m for many stars but
definition of this component is greatly influenced by the assumed
Wien tail of the dust
continuum emission and the short wavelength termination of the {\it Spitzer}
spectrum. 

Finally, we also carried out the multi-Gaussian fits on the residual {\it
Spitzer} spectra of the three hydrogen deficient PNe - A 78, A 30, and IRAS
1833-2357 - studied here for comparison. The corresponding residual spectra were
obtained by subtracting the dust continuum emission, which was represented by
5-order polynomials fitted at spectral locations free from any dust or gas
feature. Interestingly, all three PNe show weak infrared features at 6.4, 7.3,
and 8.0 $\mu$m only (see Section 4). For comparison, the parameters (central
wavelength, FWHM, and integrated flux) of the three IR emission features in
H-poor PNe are listed at the bottom of Table 3. 

\section{Overview of the residual spectra}

In both the observed and the dereddened spectra, all stars show emission
features  superimposed on the smooth continuum which is fitted with one, two,
and, in rare cases, three blackbody continua. Stars may be sorted into two
groups.

The first group with large EQWs for the emission features include  three stars 
(DY Cen, V854 Cen, and HV 2671) known (presumed in the case of HV 2671, see
above) to possess relatively large amounts of hydrogen.  Their emission
features are not limited to those in the 6-10$\mu$m interval. Residual spectra
of this trio are shown in Figure 10.  These emission features are attributed to
PAHs (see Garc\'{\i}a-Hern\'andez, Rao \& Lambert 2011b for V854 Cen and DY Cen
and Clayton et al. 2011b for HV 2671). In addition, the mid-infrared features of
the C$_{60}$ molecule or proto-fullerenes  may be  detected in DY Cen and V854
Cen (Garc\'{\i}a-Hern\'andez, Rao \& Lambert 2011b, 2012) but not in HV 2671
(Clayton et al. 2011b).  The cool RCB DY Per and the warm RCB V482 Cyg also show
PAH-like emission  features at 11.3 and 12.7 $\mu$m (see Figure 10 and Section
5). 

All other RCBs fall into the second group with an emission feature from about 6
to about 11 $\mu$m (the 6-10$\mu$m complex). The shape, essentially the
wavelength extent,  of this feature is dependent on two `external'  factors. 
First, the feature's long-wavelength profile is sensitive to the correction for
the 9.7 $\mu$m interstellar silicate absorption (Section 2.1). Second, the
feature's shape  shortward of 6 microns is sensitive to the Wien tail of the
fitted  warmer dusty blackbody and, more generally, to the lack of
contemporaneous spectral information shortward of the short-wavelength
limit of the {\it Spitzer} spectra.

In order to illustrate the quasi-continuous distribution of shapes for the 6-10
$\mu$m emission feature with minimal sensitivity to the adopted reddening, we
concentrate first on the eight RCB stars in the second group with the least
reddening, i.e.,  with adopted values of E(B-V)$\leq$0.3. 
This octet do not display a common profile for the 6-10 $\mu$m feature.
Visual inspection of
the eight spectra suggests a simple classification by the relative strength and
separation of the sharp emission features at $\sim$6.3$-$6.4 to the much broader
feature at $\sim$7-10 $\mu$m. The  apparent resolution of these two features is
influenced by the strength of emission at $\sim$7 $\mu$m. In order of increasing
strength of the 7$\mu$m feature and thus decreasing minimum between the
6.3-6.4$\mu$m and 7-10$\mu$m features, the eight are, as shown in Figure 11: V
CrA, V1157 Sgr, S Aps, RS Tel, VZ Sgr, WX CrA, U Aqr, and Z Umi. This small
sample isolated by interstellar reddening, a property extrinsic to a RCB, spans
the range of the full sample from warm to cool, minority to majority and
includes the light $s$-process enriched star U Aqr. Mean parameters from the
multi-Gaussian fits for nine components are given in Table 4. 

Examination of the relative fluxes along the sequence from V CrA to Z UMi shows
not suprisingly given the classification by morphology that the leading
variation occurs for the 6.9 $\mu$m to 6.3 $\mu$m pair which runs approximately
monotonically from 0.13 for V CrA to 1.6 for Z UMi. There is a weaker and more
erratic trend for the 7.3 $\mu$m to 6.3 $\mu$m ratio from 0.3 for V CrA to 1.0
for Z UMi. The complex of features 7.7-9.6 $\mu$m shows a nearly constant flux
ratio with respect to the 6.3 $\mu$m feature with limits of 2.4 for VZ Sgr and
WX CrA and 4.3 for S Aps but an exceptional ratio of 6.3 for U Aqr. Within the
7.7-9.6 $\mu$m complex, the relative fluxes of the various components appear
approximately the same from star to star. One or two exceptions appear to result
from an unusual FHWM assigned to a component affecting the flux of that and
adjacent components. An exception may be the 9.6 $\mu$m feature for which the
ratio relative to the 7.7 $\mu$m feature runs from 0.07 for WX CrA to 1.6 for
V1157 Sgr. Except for WX CrA and V1157 Sgr, the 7.3 $\mu$m to the 7.7-9.6 $\mu$m
flux is approximately constant which may imply that the 7.3 $\mu$m feature is a
blend composed of a feature from the 6.3 $\mu$m  carrier and the 7.7-9.6 $\mu$m
carrier.  In summary, it seems plausible to attribute the emission features to
two principal sets: one contributing the 6.3 $\mu$m and the 7.7-9.6 $\mu$m
features and another the 6.9 $\mu$m feature.  

Making use of the relative strengths of these two principal sets of  emission
features, the residual spectra in Figure 11 for the least-reddened stars and
also for the more heavily reddened stars (shown in Figures 4$-$6) may be
classifed into classes $\alpha$ to $\gamma$. The IR classes ($\alpha$,
$\beta$, $\gamma$) for all RCBs in our sample are given (when possible) in the
last column of Table 1. For the octet in Figure 11, Z UMi and U Aqr are in
class $\alpha$, VZ Sgr and WX CrA in class $\alpha\beta$, S Aps and RS Tel in
class $\beta$ and V CrA and V1157 Sgr in class $\gamma$. Across the entire
sample excluding DY Cen, V854 Cen, and HV 2671, the EQW (column EQW$_{6-10\mu
m}$ in Table 1) increases systematically down the classes from $\gamma$ to
$\alpha$, i.e., the mean EQW from four $\gamma$ stars is 0.16 $\mu$m  and 0.35
$\mu$m from six $\alpha$ stars. Within a class, the EQW appears to be constant
to within the measurement errors. The EQW is independent of stellar and dust
temperature. The greater part of this increase from $\gamma$ to $\alpha$ comes
from the growth of the 6.9 $\mu$m feature. The only gross departures from the
above trend occur for VZ Sgr with an EQW about a factor of three less than the
mean for the $\alpha\beta$ class and V3795 Sgr with an EQW about a factor of
three greater than the mean for its $\alpha$ class. Both stars are discussed in
Section 5.

The sample of the more reddened RCBs gives similar results to the
lightly-reddened sample. Inspection of the residual spectra for the
reddened sample shows greater differences star-to-star at and beyond about 10
$\mu$m than for the lightly-reddened sample. For example, the
residual spectrum for UV Cas ends at 9 $\mu$m but for V3795 Sgr the
spectrum extends beyond 12 $\mu$m but this difference is not
due to inappropriate corrections for interstellar reddening (see above).  
Average measurements of the emission features for the two samples
are compared in Table 4 and show that in the mean they share the same emission
features. This level of agreement implies that appropriate
corrections for interstellar reddening have been applied. 
 
Finally, it is worth mentioning that a secondary and very weak broad emission
feature from $\sim$11.5 to 15-18 $\mu$m is also seen in the brightest mid-IR
RCBs such as UW Cen, V517 Oph, RY Sgr, and SU Tau - although there is tentative
evidence for its presence in a few other RCBs like SV Sge, ES Aql, and WX CrA
(see Figures 3$-$6). Figure 12 displays the {\it Spitzer} residual
spectra around the weak and broad $\sim$11-15 $\mu$m emission feature for RCBs
UW Cen, ES Aql, WX CrA, V517 Oph, and SU Tau. Unfortunately, the spectra of V517
Oph and SU Tau do not cover the 5-10 $\mu$m spectral region, but UW Cen with a
strongest $\sim$11.5 to 15-18 $\mu$m broad feature displays the brightest 6-10
$\mu$m dust complex. The presence of this secondary and weaker $\sim$11.5-18
$\mu$m broad emission feature in RCB infrared spectra was revealed by previous
ISO observations of the RCB stars R CrB and RY Sgr (Lambert et al. 2001).

\section{Amorphous carbonaceous solids in H-poor environments}

Emission features in the 6-10 $\mu$m region may possibly belong to a sequence of
what were once termed `unidentified infrared bands (UIRs)' between 3 and 13
$\mu$m with principal features at wavelengths of 3.3, 6.2, 7.7, 8.7, 11.3, and
12.7  $\mu$m (Tielens 2008) in spectra of many dusty objects heated by  O-rich
and C-rich stellar atmospheres but with presumably a normal H abundance. It
would surely not be surprising if UIRs from around RCBs differed from those
around normal stars. Useful collections of mid-infrared spectra for normal stars
are illustrated and discussed by Peeters et al. (2002) from {\it ISO} 6 to 9
$\mu$m spectra and Sloan et al. (2007) from {\it Spitzer/IRS} 5 to 40 $\mu$m
spectra. 

Peeters et al. describe UIRs at 6.2 $\mu$m, a blend apparently composed of
features at 7.6, 7.8, and 8.0 $\mu$m and a feature at 8.6 $\mu$m. Extensive
discussion of the central wavelengths and profiles of these UIRs led to
identification of three classes (A, B, and C).  Sloan et al. in addition to
listing UIRS at 6.2, 7.9, and 8.6 $\mu$m report UIRs beyond the 9 $\mu$m limit
of the {\it ISO} spectra at 11.3, 12.7 and 13.7 $\mu$m.    Figure 13 composed
from residual spectra constructed from Sloan et al. (2003)'s {\it ISO} spectra
shows representative residual spectra of Peeters et al.'s classes A, B, and C,
and, in addition, the {\it Spitzer} residual spectrum of the RCB V CrA. 

Comparison with our spectra of the H-poor RCBs (see, for example, Figure 11 for
the 5-10 $\mu$m interval and Figures 3$-$6 for a longer wavelength interval)
shows a generally poor correspondence between classes A, B, and C with a normal
H abundance and H-poor RCBs. Not only is the (usually) strong 11.3 $\mu$m UIR in
classes A, B, and C absent from the RCB spectra but there are differences too
across the 6-10 $\mu$m interval; specifically,  the line at 6.9 $\mu$m present
in most RCB spectra is absent from spectra of classes A, B, and C. Apart from
the absence of UIRs beyond 10 $\mu$m, the class C spectra and those of the RCB
class $\gamma$ are quite similar  -- see IRAS 13416-6243 and V CrA in Figure
13. These differences would suggest that even where there is a close wavelength
correspondence between a UIR from classes A, B, and C and the H-poor RCBs  the
carrier of the UIR may not be the same. 

The UIRs from normal H circumstellar envelopes are attributed to different
vibrational modes of PAHs (i.e., organic structures formed of benzene rings) and
their aliphatic attachments (i.e., organic structures not containing benzene
rings). Table 4 from J\"{a}ger et al. (2008) identifies modes with  wavelength
ranges from laboratory measurements of carbon soot particles. When the aromatic
hydrocarbons are large structures, vibrational frequencies will be similar for a
given mode from one structure to the next. This physical-chemical property
explains why the UIR bands appear at very similar wavelengths in a wide variety
of sources hosting different hydrocarbons in gaseous or solid form as small
particles. Neutral and ionized PAH molecules are possible carriers with the
attractive property that they  are hardy survivors in harsh environments around
and between stars.  

Most of the RCBs are extremely H-deficient so that the circumstellar molecules
and solids are expected to also be H-poor; the UIRs in these cases have been
broadly attributed to amorphous carbon, soot particles, etc.  For the few RCBs
such as DY Cen and V854 Cen with less extreme H-deficiences, the infrared
emission features present in their {\it Spitzer} spectra have been attributed to
PAHs and (proto-)fullerenes (Garc\'{\i}a-Hern\'andez, Rao \& Lambert 2011b,
2012)\footnote{V854 Cen showed extraordinary spectral variations between the
{\it ISO} and {\it Spitzer} spectra, suggesting that a significant fraction of
the circumstellar dust grains may have evolved from hydrogenated amorphous
carbon (HAC) to PAHs and fullerene-related molecules (Garc\'{\i}a-Hern\'andez,
Rao \& Lambert 2011b; see also Garc\'{\i}a-Hern\'andez et al. 2010b, 2011,
2012).}.

Laboratory studies on PAHs and small carbon particles (amorphous carbon,
hydrogenated amorphous carbon - HAC) have characterized the band wavelengths 
and strengths.  For example, Colangeli et al. (1995)  provide wavelengths of
bands from submicron amorphous carbon particles as well as mass extinction
coefficients from 40 nm to 2 mm. Amorphous carbon was produced in three
different ways (arc discharge with carbon electrodes in an Ar [ACAR] or a H$_2$
atmosphere [ACH2], and the burning of benzene in air [BE]). The mean central
wavelengths from these experiments (ACAR, ACH2, and BE in Table 5) agree well
with the values from the multi-Gaussian fits to the spectra of the eight
least-reddened stars -- see Table 5. The 8.6, 9.1, and 9.6 $\mu$m  features were
not explicitly isolated by Colangeli et al. but illustrated emission profiles
show emission extending to these wavelengths. All Colangeli et al. laboratory
samples (ACAR, ACH2, and BE) also display a few much weaker features in the
$\sim$11-17 $\mu$m interval. In particular, the ACAR sample displays a rather
broad feature from $\sim$11.5 to $\sim$16 $\mu$m (Fig. 5 in Colangeli et al.
1995) that may be related with the secondary and weak broad $\sim$11.5-15-18
$\mu$m emission feature seen in the few brightest mid-IR RCBs (see below and
also Lambert et al. 2001). Perfect agreement between laboratory and
circumstellar wavelengths should not be expected because the precise central
wavelengths are sensitive to the nature of the amorphous carbon and, perhaps,
the circumstellar environment's  history and properties. But the wavelength
agreement does show, as expected, that large organic molecules or small carbon
grains are likely carriers of the emission features and, therefore, are also
likely responsible for the continuum infrared excess on which these features are
superimposed. Remarkably, laboratory HAC samples with very low hydrogen content
also show discrete dust features at $\sim$5.9, 6.3, 6.9, and 7.3 $\mu$m together
with the lack of emission at 11.3 $\mu$m (W. Duley, private communication).
However, the HAC laboratory samples with little H do not show any strong feature
around 8.1 $\mu$m. 

Carpentier et al. (2012) have recently studied the nanostructures of laboratory
analogues of carbonaceous circumtellar/interstellar dust, finding that the
positions of the $\sim$6.2 and 8 $\mu$m bands trace defects of the soot
polyaromatic structures in the form of non-hexagonal rings and/or aliphatic
bridges. Their sample 3 (that with the lowest hydrogenation of the carbon
skeleton) laboratory spectrum (see Figure 5 in Carpentier et al. 2012) looks
quite similar to some RCB spectra (e.g., U Aqr, V739 Sgr) where the $\sim$8
$\mu$m feature is stronger than the 6.3 $\mu$m one. Sample 3 in Carpentier et
al. (2012) provides a good laboratory analog for the carriers of the `class C'
spectra as described by Peeters et al. (2002) -- see Figure 13.  Other weaker
features at $\sim$5.8, 6.9, 7.3, and 8.6 $\mu$m - which could find a counterpart
in the RCB spectra - are also present in their sample 3 laboratory spectrum.
Insterestingly, their sample 3 is composed of big (about 30 nm in diameter) and
rather spherical primary soot particles and  characterized by a more evident
curvature of the polyaromatic structures where numerous fullerene-like
structures are found. Similar nanostructures could thus be present in the
carbonaceous dust around RCB stars. 

Table 6 presents the assignments for the infrared bands from 5 to 14 $\mu$m.
Bands between 6 and 9 $\mu$m are a mix of C-O and C-C stretching (skeletal)
modes and C-H in-plane bending modes. The 11.3 and 12.7 $\mu$m bands are  C-H
out-of-the-plane bending modes and with hydrogen being underabundant in RCB
circumstellar  shells, the absence of these bands in the majority of RCB
spectra is not surprising. The bands appear weakly in V854 Cen, DY Cen, and HV
2671, the trio with less extreme deficiencies of H. Similarly, a 3.3 $\mu$m
band composed of C-H stretching modes is absent from R CrB and RY Sgr (Lambert
et al. 2001) weakly present in V348 Sgr (Nandy, Rao \& Morgan 1986) and more
prominent in the more H-rich V854 Cen (Lambert et al. 2001).  

This explanation for the weakness of the 11.3 and 12.7 $\mu$m bands does not
simply carry over to the C-H deformation modes at about 6.9 and 7.3 $\mu$m.
Consider these C-H deformation bands at 6.9 and 7.3 $\mu$m and the reference
band at 6.3 $\mu$m arising from the C-C stretch. A naive expectation would be
that the C-H deformation band fluxes relative to that of the C-C stretch at 6.3
$\mu$m should correlate with the spectroscopically measured H abundance (Asplund
et al. 2000). In particular, one might compare H abundances for class $\alpha$
stars and class $\gamma$ stars where the latter have the  weaker 6.9-7.3$\mu$m
component. One difficulty is that H abundances are available for just two of the
four $\gamma$ stars and for four of the seven $\alpha$ stars. For the $\alpha$
stars, the logarithmic H abundances are $<$4.1, 4.8, 6.1, and 6.5. For the
$\gamma$ pair, the H abundances are 6.0 and 8.0 where the latter H abundance,
the strongest across the entire sample, applies to V CrA, the star with the
weakest 6.9-7.3$\mu$m feature. (In normal stars, the logarithmic H abundane is
12.0 by convention.) An idea that a high H abundance in the stellar atmosphere
might result in a strong flux for the 6.9 $\mu$m band is contradicted by the
appearance of V CrA's residual spectrum in Figure 11 showing a deep minimum
between the 6.3 $\mu$m band and the 8-10 $\mu$m complex just where the 6.9$\mu$m
C-H deformation should be; yet this RCB has the highest H abundance of the octet
illustrated in the Figure and most closely resembles class C spectra for stars
having a normal H abundance (see also Fig. 13).  Moreover, relative fluxes for
the two C-H deformation bands within the 6.9-7.3$\mu$m feature are uncorrelated;
the flux ratio of the 7.3 $\mu$m band to the 6.9 $\mu$m band varies over a
factor of 20 for the six stars with H abundances between 5 and 6.5. An
interpretation of this result may be that the 7.3 $\mu$m band is not entirely
due to C-H deformation. This conclusion may extend to the 6.9 $\mu$m band. There
is tentative evidence that the relative strength of the 6.9-7.3 $\mu$m feature
in RCBs may be anticorrelated with the silicon abundance. Figure 14 displays the
flux ratio 6.3$\mu$m/6.9-7.3$\mu$m versus log$\epsilon$(Si/Fe) as taken from
Asplund et al. (2000). The `minority' RCBs VZ Sgr and V CrA, stars with high
Si/Fe abundances, display the largest 6.3$\mu$m/6.9-7.3$\mu$m flux ratios.
However, the small number of stars in Figure 14 caution us about reaching a
conclusive result. In addition, the other `minority' RCB V3795 Sgr shows a
rather normal 6.9-7.3$\mu$m feature (Fig. 14).

Interestingly, there exists a clear difference in the central wavelength of the
6.3 $\mu$m feature in RCB stars with mild H-deficiencies (DY Cen, V854 Cen, 
and possibly HV 2671, and V482 Cyg\footnote{V482 Cyg is extremely H-poor
(Asplund et al. 2000) but its {\it Spitzer} spectrum display PAH-like features
and the central wavelength of its 6.3 $\mu$m feature is similar to the three
least H-deficient RCBs (see Section 5 for more details).}) and in extremely
H-deficient RCBs. The mean central wavelength is 6.26$\pm$0.02 (4 stars) and
6.36$\pm$0.03 (19 stars) for the mildly H-poor and extremely H-poor RCBs,
respectively. The central wavelenght of this feature is red-shifted up to
6.41$\pm$0.01 in the three H-deficient PNe. Similar FWHMs (within the estimated
errors) are found among the three types of objects mentioned above. However, the
6.3 $\mu$m feature in PNe with normal H abundances and compact H II regions is
blue-shifted to 6.22$\pm$0.01 (from 13 objects in Peeters et al. 2002) and it is
generally narrower (FWHM=0.13$\pm$0.02). This is shown in Figure 15, which
clearly shows the different central wavelenghts of the 6.3 $\mu$m feature
depending on the object type. Note also that the 6.3 $\mu$m feature in the RCBs
V348 Sgr, RT Nor, and Z Umi is similar to that of the H-deficient PNe (Fig. 15).
Harrington et al. (1998) reported the detection of a 6.4 $\mu$m feature in the
H-poor PN A 78 and that they identified with the 6.4 $\mu$m aromatic C-C stretch
feature produced by small, H-free carbonaceous grains\footnote{Note that
the FF star V605 Aql has also been observed with {\it Spitzer} but it shows a
featureless continuum with no silicate or PAH features (see e.g., Evans et al.
2006; Clayton et al. 2013).}. Their low quality spectrum taken with the ISOPHOT
low-resolution spectrometer did not reveal any other feature in the spectral
region from 6 to 12 $\mu$m. The higher quality residual {\it Spitzer} spectrum
of the H-poor PN A 78 displays clear features at 6.4, 7.3, and 8.0 $\mu$m. The
same set of infrared features is also present in the other H-poor PNe A 30 and
IRAS 1833$-$2357 (see Table 3). Residual {\it Spitzer} spectra of V CrA, S Aps,
and V348 Sgr - as representative examples of `minority', cool, and hot RCBs,
respectively - are compared with that of PN A 30 in Figure 16. The feature to
continuum ratio in H-poor PNe is even lower than in RCBs but the spectral
coincidence between H-poor PNe and RCB stars suggests a similar nature and
chemical composition of the dust in both circumstellar environments.

A more general search for correlations between properties (integrated flux,
central wavelength, and FWHM) of the RCB dust features (at 5.9, 6.9, 7.3, 7.7,
8.1, 8.6, 9.1, and 9.6 $\mu$m) and stellar (effective temperature, chemical
composition) and circumstellar (blackbody temperatures, filling factor) proved
negative. A variety of reasons may be suggested for this disappointment. For
example, a band's properties may be dependent on the carrier's past exposure to
photons and particles and not just to present physical conditions in the
circumstellar shell. Optical depth effects may be present. Definition of a
band's properties are subject to systematic errors which differ from
star-to-star; for example, the 5.9 $\mu$m band is sensitive to subtraction of
the underlying blackbody's Wien tail, and many bands are dependent on how other
bands, particularly adjacent bands, are fitted in the decomposition of the
residual spectrum into multi-Gaussians.  Yet, the wavelength agreement (Table 5)
between laboratory and circumstellar bands is evidence that structures 
similar to amorphous carbon reside in most RCB circumstellar shells.

\section{Exceptional infrared emission spectra?}

Inspection of the complete suite of residual spectra shows that all
but a minority may be placed in one of the classes $\alpha$ to $\gamma$.  This
minority is discussed in this section.

{\bf V3795 Sgr:}  Examination of the suite of residual spectra shows
that the 6-10 $\mu$m emission feature generally vanishes at or about 10 $\mu$m. A
clear exception is shown by V3795 Sgr (Fig. 6). Figure 7 illustrates the
residual spectra for four values of the interstellar E(B-V) including
E(B-V)=0.0 and the value E(B-V)=0.79 recommended in Paper I (Tisserand 2012 gives
E(B-V)=0.83). With the continuum fit limited to two blackbodies, the
residual spectra extend clearly to 15 $\mu$m with the long wavelength
limit declining with increasing reddening but emission longer than 10 $\mu$m cannot
be eradicated by a reasonable increase of E(B-V).  The shape of the 
residual spectrum depends sensitively on the temperature of the warmer of the two
blackbodies ($T_{BB1} = 610$ K -- Table 1) and modification of this temperature or
introduction of a non-Planckian spectrum could alter the shape of the
residual spectrum beyond 10 $\mu$m.  The extension to 15$\mu$m also
provides for an above average EQW for this $\alpha$ class RCB star.

{\bf VZ Sgr:} This star stands out in Table 1 for the EQW of the 6-10 $\mu$m
emission feature which is a factor of four smaller than for other stars in the
$\alpha\beta$ class. This fourfold difference in strength is possible tied to
the fact that VZ Sgr was about four magnitudes in V below maximum light.

{\bf MACHOJ181933:} The S/N ratio of th {\it Spitzer} spectrum is low.
 The residual spectrum is noteworthy for the lack
of emission at 7-8 $\mu$m (Fig. 5). Given the sensitivity of the emission at 5-6
$\mu$m to the Wien tail of the fitted blackbody, the weak emission in this
interval may be an artefact of the subtraction procedure, i.e.,  MACHOJ181933
may be similar to RT Nor (see below) in having a featureless residual
spectrum. Unlike RT Nor, however, MACHOJ181933's dusty shell has a covering
factor not of 1\% but of 50\% when the two blackbodies are combined (Table 1). 

{\bf DY Per:} The {\it Spitzer} observations of this cool RCB did not extend
shorter than 10 $\mu$m so that the shape of the 6-10 $\mu$m emission feature
cannot be determined. The available spectrum from 10 to 36 $\mu$m is distinctive
in that the 11.3 and 12.7 $\mu$m PAH-like features are prominent (Fig. 10).
These features occurs also in V854 Cen, DY Cen, and HV 2671, RCBs with modest H
deficiencies. This suggests that the DY Per's circumstellar shell is not
terribly H-poor, something that seems to be confirmed by our unpublishd spectra
of the CH Fraunhofer G-band.

{\bf V482 Cyg:} This RCB star is a very peculiar case. The {\it Spitzer}
residual spectrum display the 6-10 $\mu$m RCB feature but the 11.3
and 12.7 $\mu$m features that may be attributed to PAHs also appear weakly in
the spectrum of V482 Cyg (Fig. 10). According to Asplund et al (2000), this RCB
star is supposed to be extremely H-deficient and detection of 11.3 and 12.7
$\mu$m emission come us as total surprise. Interestingly, the characteristics
(i.e., the central wavelength) of its 6.3 $\mu$m  feature are similar to the
other RCB stars with normal hydrogen abundances such as DY Cen, V854 Cen, and HV
2671. V482 Cyg is known to have a K5III companion only 6" away (e.g., Rao \&
Lambert 1993). However, the {\it Spitzer} SL aperture aperture (where the 11.3
and 12.7 $\mu$m PAH-like features are seen) is only 3.7". Thus, contamination by
the nearby companion can be excluded.

{\bf RT Nor:} This RCB is one of the few RCBs with a weak infrared excess: i.e.,
$R_{BB1}$ = 0.01 in Table 1. Given the low S/N ratio of the {\it Spitzer}
spectrum and the weak infrared emission, it is not surprising that the residual
spectrum is ill-defined. The blackbody fits are not so good as for the other
RCBs and the real continuum may not be a composite of blackbody spectra. The
6-10$\mu$m emission is peculiar, peaking at  7.3 $\mu$m and there is an apparent
14-18 $\mu$m broad emission in the resulting residual spectrum (Fig. 4), but one
may draw in a non-Planckian continuum relative to which the residual spectrum
vanishes. RT Nor is not alone in our sample in having a low value of $R_{BB1}$.
Y Mus also has $R_{BB1}$ = 0.01. The residual spectrum for Y Mus (Fig. 5) is
also peaked at about 7.3 $\mu$m and there is no obvious significant emission
from 14 to 18 $\mu$m.

{\bf MV Sgr:} This is an exceptional star in our sample in that the excess
emission peaks at about 25 $\mu$m corresponding to a blackbody of about 200 K.
MV Sgr has not had a decline since at least 1986 so that this fact may
account for the presence of much cooler dust in this star. This excess declines
towards shorter wavelengths but we still see the   signature of the 6-10$\mu$m
RCB dust complex. Ground-based photometry at {\it JHKLMN} (see Paper I) show an
excess over the stellar flux for a hot (15400 K) star indicating the presence of
hot (1500-1600 K) dust in addition to the 200 K dust, both with sizeable
covering factors ($R_{BB1}$ = 0.33 for the 1500 K dust and $R_{BB2}$ = 0.18 for
the 200 K dust). The broad 6-10$\mu$m emission feature coincides with the
interval where the warm and cold dust contribute about equally to the
circumstellar flux. Changes in one or both of the blackbody contributions affect
the definition (e.g., the long-wavelength end) of the emission feature, as would
the introduction of flux from dust at an intermediate temperature. The MV Sgr
6-10$\mu$m emission shows the peculiarity of the 7.3 $\mu$m feature being the
most intense feature. This contrasts with the case of V348 Sgr, another hot RCB
where the 8.1 $\mu$m feature dominates in the 6-10 $\mu$m interval.

{\bf V348 Sgr:} Although V348 Sgr is among the least H-poor of RCBs (Jeffery
1995)  and of a similar effective temperature to similarly `H-rich' RCBs
as DY Cen and HV 2671, its residual spectrum is unlike
the latter pair and resembles the spectra of H-poor RCBs.

\section{Conclusions}

We have presented and discussed residual {\it Spitzer/IRS} spectra for a sample
of 31 RCB stars, including warm RCB stars across the chemical composition range
observed, the coolest RCB stars, and minority RCBs. The RCB dust features
present in the residual spectra do not strongly depend on the adopted
interstellar   reddening as shown by our distinction between the least reddening
and more reddening RCB stars. 

A broad $\sim$6-10 $\mu$m dust emission feature is detected  in the residual
spectra of almost all RCB stars with extreme H-deficiencies. In the few
brightest RCBs, the $\sim$6-10 $\mu$m dust emission complex is accompanied by a
secondary and weaker broad $\sim$11.5-15 $\mu$m emission feature. These broad
features were previously seen in the lower S/N ISO spectra of R CrB and RY Sgr
but the {\it Spitzer/IRS} observations presented here reveal the detailed
structure within the broad $\sim$6-10 $\mu$m dust emission complex for the first
time. Depending on the relative strenghts of two principal sets of emission
features (the 6.3$\mu$m and 7.7-9.6$\mu$m features and the 6.9$\mu$m feature),
extremely H-poor RCB stars may be classified in several classes from $\alpha$ to
$\gamma$. The few least H-deficient RCBs except for V348 Sgr, however, display
much richer infrared spectra, with the presence of narrower features that may be
attributed to PAHs and fullerene-related molecules. 

By using the highest-quality RCB residual spectra (e.g., UW Cen) as a
reference, we find that the $\sim$6-10 $\mu$m RCB complex is resolvable into
indivivual emission features at $\sim$5.9, 6.3, 6.9, 7.3, 7.7, 8.1, 8.6, 9.1,
and 9.6 $\mu$m. A multi-Gaussian fit across the RCB sample shows that these
infrared emission features have similar wavelengths and widths. Interestingly,
we find a reasonably good agreement between the circumstellar wavelengths and
those measured in the laboratory for small carbon particles such as amorphous carbon
and HAC grains with little H. We conclude that infrared emission features in the
H-poor environments of RCBs are consistent with the carriers being amorphous
carbonaceous solids with little or no hydrogen. In addition, the spectral
coincidence between RCB stars and H-poor PNe suggests a similar nature and
chemical composition of the dust in both circumstellar environments.

Remarkably, the wavelength position of the 6.3 $\mu$m emission feature is found
to be dependent on the object type; mildly H-deficient and extremely H-deficient
RCBs and H-poor PNe. However, our general search for correlations between the
characteristics of the RCB infrared emission features (central wavelength,
integrated flux, FWHM) and stellar (effective temperature, chemical composition)
and circumstellar (blackbody dust temperature, filling factor) properties has
been proved negative. This possibly suggests that the RCB dust features are
sensitive to the chemical composition and thermal/photochemical history of the
amorphous carbon solids in the circumstellar shells rather than to the actual
physical conditions around RCBs. 


%% Included in this acknowledgments section are examples of the
%% AASTeX hypertext markup commands. Use \url without the optional [HREF]
%% argument when you want to print the url directly in the text. Otherwise,
%% use either \url or \anchor, with the HREF as the first argument and the
%% text to be printed in the second.

\acknowledgments

This work is based on observations made with the {\it Spitzer Space Telescope},
which is operated by the Jet Propulsion Laboratory, California Institute of
Technology, under NASA contract 1407. We acknowledge the referee Geoff
Clayton for comments that helped to improve the paper. We also acknowledge W.
Duley for providing us with his unpublished laboratory data of HAC samples.
D.A.G.H. acknowledges support for this work provided by the Spanish Ministry of
Economy and Competitiveness under grant AYA$-$2011$-$27754. D.L.L. acknowledges
support for this work provided by NASA through an award for program GO \#50212
issued by JPL/Caltech and  wishes to thank the Robert A. Welch Foundation of
Houston, Texas for support through grant F-634. N.K.R. thanks A.V. Raveendran
and P. Ramya for assistance with the Gaussian-fitting routine and David and
Melody Lambert for hospitality in Austin where part of this work was done.


%% in parentheses, after the keyword, but they will not be verified.

{\it Facilities:} \facility{Spitzer (IRS)}.

\clearpage

\begin{deluxetable}{lrllrrclll}
\tabletypesize{\scriptsize}
\tablecaption{Properties of the RCB stars sample \label{tbl-1}}
\tablewidth{0pt}
\tablehead{
\colhead{RCB star} & \colhead{T$_{star}$} & \colhead{T$_{BB1}$} & 
 \colhead{R$_{BB1}$} & \colhead{T$_{BB2}$} & \colhead{R$_{BB2}$}
&  \colhead{EQW$_{6-10\mu m}$} & \colhead{Var.$^{a}$} & 
\colhead{E(B-V)$^{b}$} & \colhead{IR Class}\\
\hline
\hline
 \colhead{} & \colhead{(K)} & \colhead{(K)} & \colhead{} & \colhead{(K)} & \colhead{} 
 & \colhead{($\mu$m)} & \colhead{} & \colhead{} & \colhead{}
 }
\startdata
 RY Sgr        &7200   & 675 & 0.20  &$\dots$ &$\dots$ & 0.26$^{c}$ & min& 0.00 & $\dots$       \\
 R CrB         &6750   & 950 & 0.30  &$\dots$ &$\dots$ & 0.88$^{c}$ & max& 0.00 & $\dots$       \\
 Z UMi         &5200   & 710 & 0.43  &$\dots$ &$\dots$ & 0.32       & min& 0.00 & $\alpha$      \\
 S Aps         &4200   & 730 & 0.37  &$\dots$ &$\dots$ & 0.27       & max& 0.05 & $\beta$       \\
 U Aqr         &5000   & 473 & 0.23  & 140    & 0.021  & 0.29       & max& 0.05 & $\alpha$      \\
 WX CrA        &4200   & 575 & 0.15  & 120    & 0.006  & 0.25       & max& 0.06 & $\alpha\beta$ \\
 V854 Cen      &6750   & 900 & 0.32  & 140    & 0.030  & 0.87$^{d}$ & min& 0.07 & $\alpha$      \\
 V CrA         &6500   & 552 & 0.38  & 156    & 0.020  & 0.11       & max& 0.14 & $\gamma$      \\
 HV2671        &20000  & 590 & 0.36  & 150    & 0.268  & 0.76$^{e}$ &$\dots$ & 0.15 & $\dots$   \\
 RS Tel        &6750   & 830 & 0.25  & 135    & 0.005  & 0.21       & max& 0.17 & $\beta$       \\
 V1157 Sgr     &4200   & 770 & 0.59  & 120    & 0.007  & 0.19       & min& 0.30 & $\gamma$      \\
 VZ Sgr        &7000   & 750 & 0.17  & 140    & 0.008  & 0.09       & min& 0.30 & $\alpha\beta$ \\
 UW Cen        &7500   & 636 & 0.44  & 120    & 0.013  & 0.21       & max& 0.32 & $\alpha$      \\
 ES Aql        &4500   & 774 & 0.49  &$\dots$ &$\dots$ & 0.18       & max& 0.32 & $\beta$       \\
 RT Nor        &6700   & 365 & 0.01  & 152    & 0.001  & 0.20       & max& 0.39 & $\dots$       \\
 V1783 Sgr     &5600   & 560 & 0.28  &$\dots$ &$\dots$ & 0.50       & max& 0.42 & $\alpha$      \\
 MV Sgr        &15400  &1640 & 0.32  & 207    & 0.160  & 0.90       & max& 0.43 & $\dots$       \\
 V348 Sgr      &20000  & 707 & 0.63  & 100    & 0.035  & 0.36$^{f}$ & min& 0.45 & $\alpha$      \\
 DY Cen        &19500  & 272 & 0.09  &$\dots$ &$\dots$ & 2.76$^{g}$ & max& 0.47 & $\dots$       \\
 DY Per        &3000   &1400 & 0.31  &$\dots$ &$\dots$ &$\dots$     & max& 0.48 & $\dots$       \\
 V517 Oph      &4100   & 886 & 0.84  &$\dots$ &$\dots$ &$\dots$     & min& 0.50 & $\dots$       \\
 SU Tau        &6500   & 637 & 0.45  &$\dots$ &$\dots$ &$\dots$	    & max& 0.50 & $\dots$       \\
 Y Mus         &7200   & 395 & 0.01  &$\dots$ &$\dots$ & 0.38       & max& 0.50 & $\alpha$      \\
 V739 Sgr      &5400   & 640 & 0.59  & 100    & 0.005  & 0.18       & max& 0.50 & $\beta$       \\
 MACHOJ181933  &4200   &710  & 0.48  & 130    & 0.022  & 0.03       & max& 0.50 & $\dots$       \\
 RZ Nor        &5000   & 698 & 0.53  & 325    & 0.035  & 0.44       & max& 0.50 & $\beta$       \\
 V482 Cyg      &4800   & 590 & 0.03  & 150    & 0.001  & 0.39       & max& 0.50 & $\alpha$      \\
 SV Sge        &4200   & 475 & 0.05  & 370    & 0.024  & 0.50       & max& 0.83 & $\alpha\beta$ \\
 V3795 Sgr     &8000   & 610 & 0.31  &$\dots$ &$\dots$ & 1.1        & max& 0.79 & $\alpha$      \\
 UV Cas        &7200   & 510 & 0.03  & 180    & 0.001  & 0.15       & max& 0.90 & $\beta$       \\
 FH Sct        &6250   & 540 & 0.10  & 135    & 0.002  & 0.25       & max& 1.00 & $\beta$       \\
\enddata
\tablenotetext{a}{Variability status during the {\it Spitzer} observations.}
\tablenotetext{b}{See Paper I for more details about the adopted E(B-V) values.}
\tablenotetext{c}{Estimated from ISO spectrum (Lambert et al. 2001).}
\tablenotetext{d}{Equivalent width from features between 5.7 and 10.4 $\mu$m.}
\tablenotetext{e}{Equivalent width from features between 5.6 and 9.5 $\mu$m.}
\tablenotetext{f}{Equivalent width from features between 5.8 and 10.8 $\mu$m.}
\tablenotetext{g}{Equivalent width from features between 5.6 and 12.4 $\mu$m.}
\end{deluxetable}

\clearpage

\begin{landscape}

\begin{deluxetable}{lccccccccc}
\tabletypesize{\scriptsize}
\tablecaption{Dust features for the low E(B-V) ($\leq$0.30) RCB stars$^{a}$\label{tbl-1}}
\tablewidth{0pt}
\tablehead{
\colhead{RCB star} & \colhead{5.9$\mu$m} & \colhead{6.3$\mu$m} & \colhead{6.9$\mu$m} &
\colhead{7.3$\mu$m} & \colhead{7.7$\mu$m} & \colhead{8.1$\mu$m} & \colhead{8.6$\mu$m} &
\colhead{9.1$\mu$m} & \colhead{9.6$\mu$m} \\
\hline
\hline
\colhead{}  & \colhead{$\lambda$[F, FWHM]} & \colhead{$\lambda$[F, FWHM]} & \colhead{$\lambda$[F, FWHM]} &
\colhead{$\lambda$[F, FWHM]} & \colhead{$\lambda$[F, FWHM]} & \colhead{$\lambda$[F, FWHM]} & \colhead{$\lambda$[F, FWHM]} &
\colhead{$\lambda$[F, FWHM]} & \colhead{$\lambda$[F, FWHM]} \\
\hline
\colhead{}  & \colhead{($\mu$m[$^{b}$, $\mu$m])} & \colhead{($\mu$m[$^{b}$, $\mu$m])} &
\colhead{($\mu$m[$^{b}$, $\mu$m])} & \colhead{($\mu$m[$^{b}$, $\mu$m])} & \colhead{($\mu$m[$^{b}$, $\mu$m])} &
\colhead{($\mu$m[$^{b}$, $\mu$m])} & \colhead{($\mu$m[$^{b}$, $\mu$m])} & \colhead{($\mu$m[$^{b}$, $\mu$m])} &
\colhead{($\mu$m[$^{b}$, $\mu$m])} 
}
\startdata
 Z UMi  & 5.95[2.0, 0.41] & 6.40[2.9, 0.47] &  6.86[4.5, 0.63] & 7.32[2.8, 0.50] & 7.73[4.6, 0.59] & 8.21[5.4, 0.62] & 8.67[2.4, 0.47] & 9.12[2.5, 0.58] & 9.68[1.3, 0.57] \\
 S Aps  & 5.83[1.6, 0.34] & 6.39[7.1, 0.42] &  6.90[4.7, 0.44] & 7.23[1.0, 0.16] & 7.68[8.6, 0.75] & 8.15[9.7, 0.85] & 8.64[6.8, 0.72] & 9.16[3.7, 0.56] & 9.70[2.0, 0.44] \\
 U Aqr  & 5.86[0.2, 0.21] & 6.34[1.2, 0.45] &  6.86[1.4, 0.67] & 7.26[0.7, 0.38] & 7.61[1.4, 0.56] & 8.11[2.4, 0.63] & 8.66[2.0, 0.65] & 9.11[0.8, 0.44] & 9.68[1.0, 0.72] \\
 WX CrA & 5.94[0.2, 0.21] & 6.37[2.4, 0.45] &  6.81[1.1, 0.40] & 7.28[3.0, 0.68] & 7.72[1.4, 0.48] & 8.11[2.2, 0.52] & 8.60[1.5, 0.50] & 9.10[0.6, 0.45] & 9.57[0.1, 0.51] \\
  V CrA & 5.96[0.6, 0.30] & 6.33[2.3, 0.35] &  6.83[0.3, 0.32] & 7.23[0.6, 0.25] & 7.70[2.2, 0.45] & 8.14[3.4, 0.50] & 8.63[3.7, 0.58] & 9.13[1.6, 0.59] & 9.75[1.2, 0.75] \\
 RS Tel & 5.90[0.3, 0.22] & 6.36[2.7, 0.37] &  6.86[2.3, 0.48] & 7.26[1.4, 0.35] & 7.73[3.6, 0.62] & 8.12[1.9, 0.52] & 8.49[2.5, 0.65] & 9.08[1.1, 0.47] & 9.62[0.2, 0.38] \\
V1157Sgr& 5.57[0.6, 0.31] & 6.39[3.7, 0.55] &  6.90[1.0, 0.45] & 7.29[1.5, 0.31] & 7.65[2.1, 0.40] & 8.15[4.8, 0.58] & 8.69[2.9, 0.43] & 9.10[1.6, 0.34] & 9.55[3.3, 0.62] \\
 VZ Sgr & $\dots$	  & 6.36[1.6, 0.39] &  6.86[0.9, 0.51] & 7.19[0.5, 0.50] & 7.54[0.7, 0.45] & 8.11[1.4, 0.67] & 8.72[0.8, 0.61] & 9.18[0.4, 0.55] & 9.65[0.6, 0.55] \\
\enddata
\tablenotetext{a}{Maximum flux errors in RCBs are estimated to be of the order of $\sim$30$-$40 \%.} 
\tablenotetext{b}{The listed fluxes (F) are in units of 10$^{-19}$ W cm$^{-2}$.} 
\end{deluxetable}

\end{landscape}

\clearpage

\begin{landscape}

\begin{deluxetable}{lccccccccc}
\tabletypesize{\scriptsize}
\tablecaption{Dust features for the high E(B-V) ($>$0.30) RCB stars$^{a,b}$\label{tbl-1}}
\tablewidth{0pt}
\tablehead{
\colhead{RCB star} & \colhead{5.9$\mu$m} & \colhead{6.3$\mu$m} & \colhead{6.9$\mu$m} &
\colhead{7.3$\mu$m} & \colhead{7.7$\mu$m} & \colhead{8.1$\mu$m} & \colhead{8.6$\mu$m} &
\colhead{9.1$\mu$m} & \colhead{9.6$\mu$m} \\
\hline
\hline
\colhead{}  & \colhead{$\lambda$[F, FWHM]} & \colhead{$\lambda$[F, FWHM]} & \colhead{$\lambda$[F, FWHM]} &
\colhead{$\lambda$[F, FWHM]} & \colhead{$\lambda$[F, FWHM]} & \colhead{$\lambda$[F, FWHM]} & \colhead{$\lambda$[F, FWHM]} &
\colhead{$\lambda$[F, FWHM]} & \colhead{$\lambda$[F, FWHM]} \\
\hline
\colhead{}  & \colhead{($\mu$m[$^{c}$, $\mu$m])} & \colhead{($\mu$m[$^{c}$, $\mu$m])} &
\colhead{($\mu$m[$^{c}$, $\mu$m])} & \colhead{($\mu$m[$^{c}$, $\mu$m])} & \colhead{($\mu$m[$^{c}$, $\mu$m])} &
\colhead{($\mu$m[$^{c}$, $\mu$m])} & \colhead{($\mu$m[$^{c}$, $\mu$m])} & \colhead{($\mu$m[$^{c}$, $\mu$m])} &
\colhead{($\mu$m[$^{c}$, $\mu$m])} 
}
\startdata
 UW Cen & 5.94[11.1, 0.39] & 6.38[10.0, 0.39] &  6.70[3.7, 0.23] & 7.23[17.0, 0.70] & 7.69[5.6, 0.36] & 8.06[23.5, 0.63] & 8.62[14.0, 0.59] & 9.12[0.3, 0.27] & 9.50[1.5, 0.35] \\
 ES Aql & 5.97[0.5, 0.20]  & 6.36[3.2, 0.49]  &  6.85[1.5, 0.38] & 7.24[0.9, 0.25]  & 7.65[3.2, 0.50] & 8.10[3.9, 0.54]  & 8.63[2.8, 0.58]  & 9.13[0.6, 0.33] & 9.54[0.6, 0.49] \\
 RT Nor & 5.96[0.03, 0.22] & 6.41[0.3, 0.37]  &  6.86[0.3, 0.48] & 7.29[0.3, 0.35]  & 7.73[0.09, 0.42]& 8.18[0.09, 0.52] & 8.49[0.1, 0.65]  & 9.08[0.1, 0.47] & 9.62[0.05, 0.38] \\
V1783Sgr& 5.91[2.6, 0.30]  & 6.37[10.6, 0.60] &  6.90[7.0, 0.50] & 7.27[5.5, 0.40]  & 7.67[9.7, 0.54] & 8.27[17.2, 0.79] & 8.89[7.5, 0.64]  & 9.40[5.1, 0.70] & 9.96[6.0, 0.81] \\
 MV Sgr & 5.97[3.3, 0.28]  & 6.38[1.3, 0.51]  &  6.84[1.1, 0.46] & 7.26[1.3, 0.45]  & 7.70[1.4, 0.55] & 8.11[1.1, 0.55]  & 8.58[1.6, 0.63]  & 9.10[1.2, 0.57] & 9.66[1.3, 0.58] \\
 V348Sgr& 5.95[4.7, 0.38]  & 6.41[6.5, 0.41]  &  6.88[3.2, 0.46] & 7.28[4.4, 0.40]  & 7.70[10.5, 0.55]& 8.11[11.0, 0.52] & 8.58[13.8, 0.65] & 9.15[11.0, 0.67]& 9.79[11.1, 0.85] \\
  Y Mus & $\dots$	   & 6.35[0.6, 0.48]  &  6.80[0.4, 0.35] & 7.28[0.7, 0.44]  & 7.70[0.5, 0.43] & 8.12[0.4, 0.48]  & 8.63[0.4, 0.50]  & 9.15[0.2, 0.34] & 9.66[0.1, 0.39] \\
 V739Sgr& 5.84[0.4, 0.25]  & 6.36[1.3, 0.35]  &  6.84[1.2, 0.50] & 7.26[0.6, 0.60]  & 7.69[1.8, 0.44] & 8.11[2.0, 0.50]  & 8.61[2.1, 0.59]  & 9.20[1.8, 0.67] & 9.82[1.2, 0.45] \\
MACHOJ18& 5.92[0.5, 0.34]  & 6.34[0.3, 0.37]  & $\dots$		 & $\dots$          & $\dots$	      & 8.11[0.1, 0.28]  & $\dots$	    & $\dots$	      & $\dots$	     \\	  
 RZ Nor & $\dots$	   & 6.33[7.0, 0.43]  &  6.80[6.5, 0.48] & 7.24[6.4, 0.44]  & 7.68[11.0, 0.57]& 8.17[8.5, 0.47]  & 8.61[8.5, 0.53]  & 9.09[6.5, 0.64] & 9.65[7.4, 0.76] \\
 V482Cyg& 5.69[0.07, 0.22] & 6.28[1.8, 0.46]  &  6.76[2.0, 0.54] & 7.24[1.8, 0.54]  & 7.77[2.3, 0.79] & 8.09[2.2, 0.79]  & 8.58[1.2, 0.54]  & 9.03[0.7, 0.53] & 9.69[0.2, 0.45] \\ 
 SV Sge & 5.78[3.1, 0.45]  & 6.37[15.8, 0.69] &  6.84[3.8, 0.40] & 7.29[13.9, 0.63] & 7.80[5.3, 0.44] & 8.16[8.1, 0.64]  & 8.88[10.0, 0.95] & $\dots$	      & 9.59[4.5, 0.63] \\
V3795Sgr& 5.60[2.2, 0.56]  & 6.38[17.3, 0.69] &  6.88[10.0, 0.82]& 7.21[15.8, 0.80] & 7.68[17.8, 0.77]& 8.16[17.5, 0.67] & 8.65[15.0, 0.66] & 9.22[19.5, 0.82]& 9.89[11.8, 0.74] \\
 UV Cas & $\dots$	   & 6.32[1.0, 0.33]  &  6.72[0.6, 0.30] & 7.27[1.3, 0.42]  & 7.67[0.9, 0.41] & 8.12[2.2, 0.58]  & 8.65[1.0, 0.47]  & 9.17[0.3, 0.42] & 9.68[0.4, 0.40] \\
 FH Sct & 5.92[4.2, 0.31]  & 6.37[2.7, 0.41]  &  6.81[1.3, 0.36] & 7.21[1.8, 0.41]  & 7.64[2.3, 0.52] & 8.10[2.5, 0.58]  & 8.61[1.9, 0.53]  & 9.20[1.3, 0.61] & 9.76[1.5, 0.63] \\
\hline
H-rich RCBs &  &  &   & &   & &  &  &  \\
V854 Cen& 5.97[15.0, 0.22]& 6.27[84.5, 0.24]  &  6.58[52.5, 0.50]& 7.02[70.0, 0.64] & 7.87[314.0, 0.85]& $\dots$ 	         & 8.51[86.0, 0.55] & 9.11[38.7, 0.53]& 9.79[14.0, 0.64] \\
 HV2671 & 5.83[0.2, 0.30] & 6.24[0.6, 0.38]   &  6.75[0.6, 0.55] & 7.29[0.5, 0.41]  & 7.65[0.9, 0.58] & 8.10[0.2, 0.40]  & 8.58[0.5, 0.56]  & 9.13[0.06, 0.32]& 9.69[0.03, 0.32] \\
 DY Cen & $\dots$	      & 6.25[3.3, 0.22]   &  6.58[3.3, 0.47] & 7.02[3.5, 0.38]  & 7.74[16.3, 0.81]& $\dots$ 	         & 8.55[4.7, 0.57]  & 9.14[1.2, 0.79] & 9.79[1.9, 0.84] \\
\hline
H-poor PNe$^{d}$ &  &  &   & &   & &  &  &  \\
A 78 & $\dots$ & 6.43[0.3, 0.32] & $\dots$  & 7.31[0.2, 0.20] & $\dots$ 	       & 7.96[0.1, 0.29] & $\dots$ & $\dots$ & $\dots$ \\
A 30 & $\dots$ & 6.40[0.4, 0.35] & $\dots$  & 7.33[0.3, 0.27] & $\dots$ 	       & 7.96[0.1, 0.32] & $\dots$ & $\dots$ & $\dots$ \\
IRAS 1833$-$2357 & $\dots$ & 6.40[0.2, 0.34] & $\dots$  & 7.29[0.2, 0.31] & $\dots$  & 7.96[0.03, 0.22]$^{e}$ & $\dots$ & $\dots$ & $\dots$ \\
\enddata
\tablenotetext{a}{Note that we also list at the bottom the three least H-deficient RCBs and the three H-poor PNe.}
\tablenotetext{b}{Maximum flux errors in RCBs are estimated to be of the order of $\sim$30$-$40 \%.} 
\tablenotetext{c}{The listed fluxes (F) are in units of 10$^{-19}$ W cm$^{-2}$.} 
\tablenotetext{d}{Flux errors are possibly higher than those for RCBs.} 
\tablenotetext{e}{The measurements are rather uncertain because the feature is weak.}
\end{deluxetable}

\end{landscape}

\clearpage

\begin{deluxetable}{lccccc}
\tabletypesize{\scriptsize}
\tablecaption{Mean properties of the dust features in RCBs \label{tbl-1}}
\tablewidth{0pt}
\tablehead{
\colhead{} & \colhead{Least-reddened RCBs$^{a}$} & \colhead{} & \colhead{} & \colhead{More-reddened RCBs$^{b}$} 
& \colhead{} \\
\hline 
\hline
\colhead{Feature} & \colhead{$\lambda$} & \colhead{FWHM} & \colhead{} & \colhead{$\lambda$} & \colhead{FWHM} \\
 \hline
 \colhead{} & \colhead{($\mu$m)} & \colhead{($\mu$m)} & \colhead{} 
 & \colhead{($\mu$m)} & \colhead{($\mu$m)}
 }
\startdata
5.9$\mu$m &  5.86$\pm$0.14 & 0.29$\pm$0.08 &  &5.87$\pm$0.12&0.33$\pm$0.11    \\
6.3$\mu$m &  6.37$\pm$0.03 & 0.43$\pm$0.06 &  &6.36$\pm$0.03&0.47$\pm$0.12    \\
6.9$\mu$m &  6.86$\pm$0.03 & 0.49$\pm$0.12 &  &6.82$\pm$0.06&0.45$\pm$0.14    \\
7.3$\mu$m &  7.26$\pm$0.04 & 0.39$\pm$0.16 &  &7.26$\pm$0.03&0.49$\pm$0.15    \\
7.7$\mu$m &  7.67$\pm$0.07 & 0.54$\pm$0.12 &  &7.70$\pm$0.04&0.52$\pm$0.13    \\
8.1$\mu$m &  8.14$\pm$0.03 & 0.61$\pm$0.11 &  &8.13$\pm$0.05&0.57$\pm$0.13    \\
8.6$\mu$m &  8.64$\pm$0.07 & 0.58$\pm$0.10 &  &8.64$\pm$0.11&0.61$\pm$0.12    \\
9.1$\mu$m &  9.12$\pm$0.03 & 0.50$\pm$0.09 &  &9.16$\pm$0.09&0.54$\pm$0.17    \\
9.6$\mu$m &  9.65$\pm$0.07 & 0.57$\pm$0.13 &  &9.70$\pm$0.13&0.57$\pm$0.17    \\
\enddata
\tablenotetext{a}{Mean values obtained for the eight least-reddened (E(B-V)$\leq$0.30) RCBs (see Fig.3).}
\tablenotetext{b}{Mean values obtained for the fifteen more-reddened (E(B-V)$>$0.30) RCBs (see Figures 4$-$6)}
\end{deluxetable}

\clearpage

\begin{deluxetable}{llllr}
\tabletypesize{\scriptsize}
\tablecaption{Central wavelengths for amorphous carbon samples and the least-reddened RCBs\label{tbl-1}}
\tablewidth{0pt}
\tablehead{
\colhead{} &\colhead{AC Samples$^{a}$} &  \colhead{} & \colhead{} & \colhead{RCBs$^{b}$} \\
\hline
\hline
\colhead{ACH2} & \colhead{ACAR} & \colhead{BE} & \colhead{Mean$_{AC}$} &
\colhead{Mean$_{RCBs}$} \\
\hline
\colhead{($\mu$m)} & \colhead{($\mu$m)} &\colhead{($\mu$m)} &\colhead{($\mu$m)} & \colhead{($\mu$m)} 
 }
\startdata
  5.84    & 5.80    & $\dots$ & 5.82    & 5.84$\pm$0.03   \\
  6.22    & 6.40    & 6.33    & 6.32    & 6.37$\pm$0.03   \\
  6.93    & 6.91    & 6.95    & 6.93    & 6.85$\pm$0.05   \\
  7.28    & 7.29    & 7.24    & 7.27    & 7.25$\pm$0.03    \\
  8.10    & 7.96    & 7.72    & 7.93    & 8.14$\pm$0.05    \\
%  $\dots$ & $\dots$ & $\dots$ & $\dots$ & 8.64$\pm$0.09    \\
\enddata
\tablenotetext{a}{From Colangeli et al. (1995) - see text.}
\tablenotetext{b}{From the multi-Gaussian fits to the residual spectra of the eight least-reddened (E(B-V)$\leq$0.30)
RCBs.}
\end{deluxetable}

\clearpage

\begin{deluxetable}{lcll}
\tabletypesize{\scriptsize}
\tablecaption{Assignments for the infrared bands from 5 to 14 $\mu$m $^{a}$\label{tbl-1}}
\tablewidth{0pt}
\tablehead{
\colhead{Band} & \colhead{Assignment} & \colhead{} & \colhead{} \\
\hline
\hline
\colhead{($\mu$m)} & \colhead{} & \colhead{} & \colhead{} 
 }
\startdata
  5.78 - 5.81 & -C=O stretch        &  \\
  6.23 - 6.28 & -C=C- stretch       & 	  \\
  6.83 - 6.87 & -C-H deformation    &    \\
  7.25 - 7.35 & -C-H deformation    &   \\
  7.81 - 7.87 & -C-C- stretch       & -C-H deformation  \\
  8.73 - 8.92 & -C-C- stretch       & -C-H deformation  \\
 11.6 - 11.8  & =C-H out of plane,   &  1 H, 2 adjacent H \\
 13.4 - 13.6  & =C-H out of plane,   &  3-5 adjacent H \\
\enddata
\tablenotetext{a}{From J\"{a}ger et al. (2008)}
\end{deluxetable}

\clearpage

%All Figures

\begin{figure}
%Figure 1
\begin{center}
\includegraphics[angle=0,scale=.40]{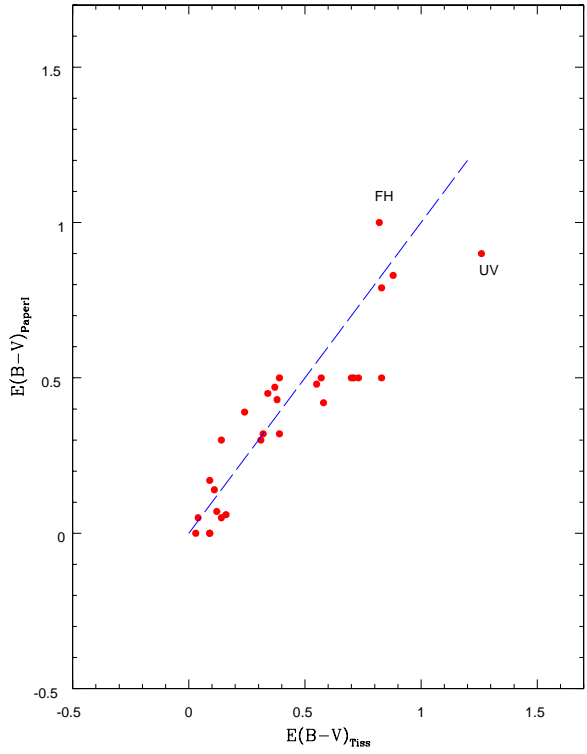} 
\end{center}
\caption{Comparison of the ISM reddening E(B-V) estimates due to Tisserand
(2012) and us (Paper I). Tisserand's estimates are obtained from COBE/DIRBE maps
(Schlegel et al. 1998) and our estimates are taken from Paper I. The dashed line
is a 45 degrees line. This comparison suggests that the E(B-V) estimates broadly
agree but for two exceptions (FH = FH Sct and UV = UV Cas).
\label{fig1}}
\end{figure} 

\clearpage

\begin{figure}
%Figure 2
\begin{center}
\includegraphics[angle=0,scale=.40]{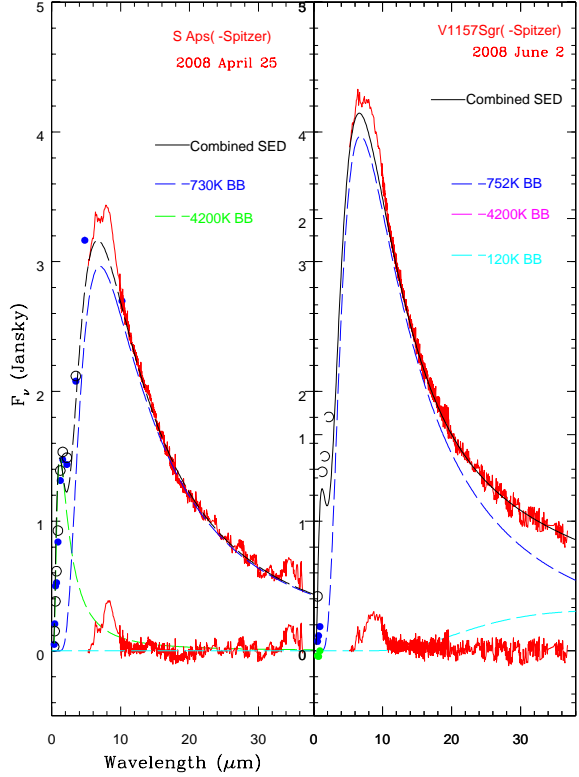} 
\end{center}
\caption{Illustrative example of the blackbody fits adopted for the RCB stars S
Aps (left panel) and V1157 Sgr (right panel). For S Aps (E(B-V)=0.05), a
reddening correction is ignored and a combination of blackbodies of 4200 K for
the star (green dashed line) and 730 K for the circumstellar dust (blue dashed
line) gives a good fit (black dashed line) to the spectral energy distribution
(SED) from $\sim$4 to 37 $\mu$m. V1157 Sgr is  mildly reddened  with a stellar
temperature (4200 K) similar to S Aps.  The reddening-corrected {\it Spitzer}
spectrum of V1157 Sgr (in red) has been obtained for E(B-V) = 0.30 using the
reddening curve from Chiar \& Tielens (2006). A combination of three blackbodies
of 4200 K (the stellar component), 752 and 120 K (the infrared circumstellar
component) provides a good fit to V1157 Sgr's SED. Note that the resulting
residual spectra for both RCBs are shown at the bottom for
comparison. \label{fig2}}
\end{figure} 

\clearpage

\begin{figure}
%Figure 3
\includegraphics[angle=0,scale=.60]{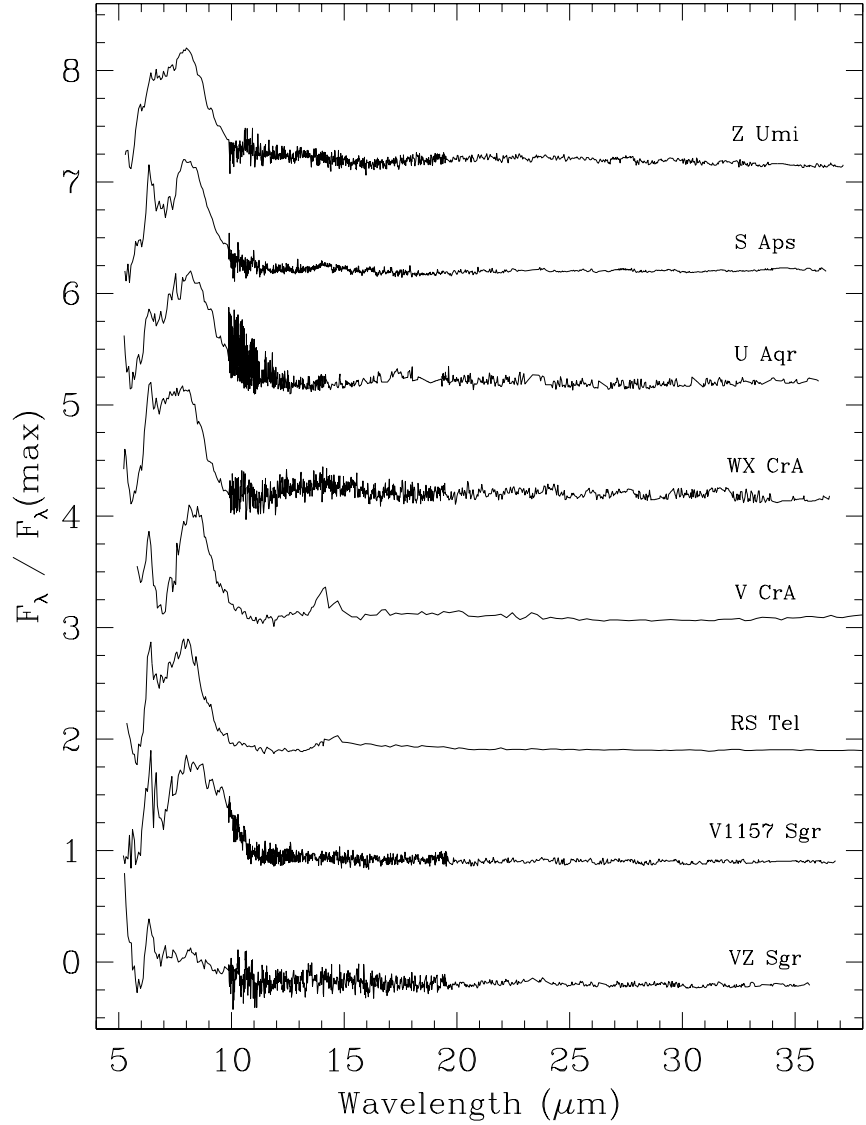}
\caption{{\it Spitzer/IRS} residual spectra over the full wavelength range
$\sim$5$-$37 $\mu$m for the eight low E(B-V) ($\leq$0.30) RCB stars. The stars
ordered by increasing E(B-V) from top to bottom are: Z UMi, S Aps, U Aqr, WX
CrA, V CrA, RS Tel, V1157 Sgr, and VZ Sgr. Note that the spectra are normalized
and displaced for clarity. \label{fig3}}
\end{figure}

\clearpage

\begin{figure}
%Figure 4
\includegraphics[angle=0,scale=.60]{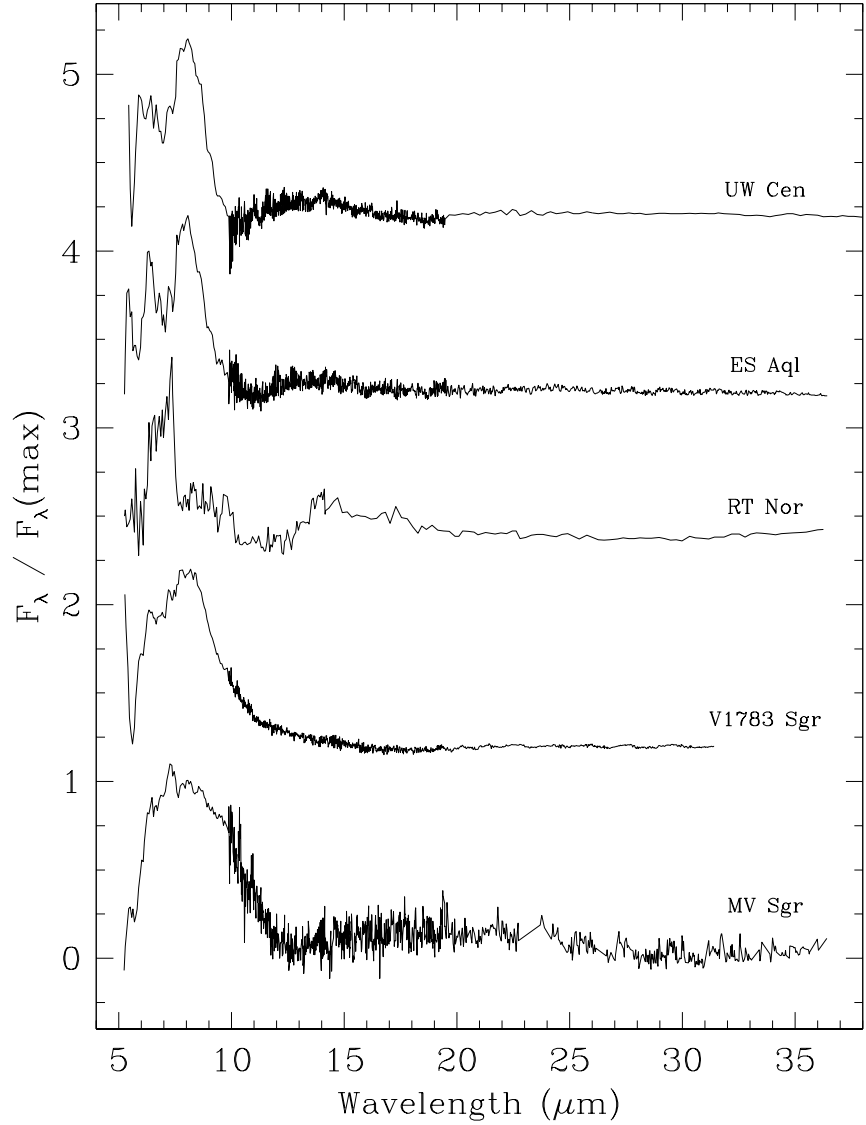}
\caption{{\it Spitzer/IRS} residual spectra over the full wavelength
range $\sim$5$-$37 $\mu$m for high E(B-V) ($>$0.30) RCB stars UW Cen, ES Aql, RT
Nor, V1783 Sgr, and MV Sgr, which are ordered top to bottom by increasing E(B-V)
from E(B-V) = 0.32 for UW Cen to 0.43 for MV Sgr.  Note that the spectra are
normalized and displaced for clarity. \label{fig4}}
\end{figure}

\clearpage

\begin{figure}
%Figure 5
\includegraphics[angle=0,scale=.60]{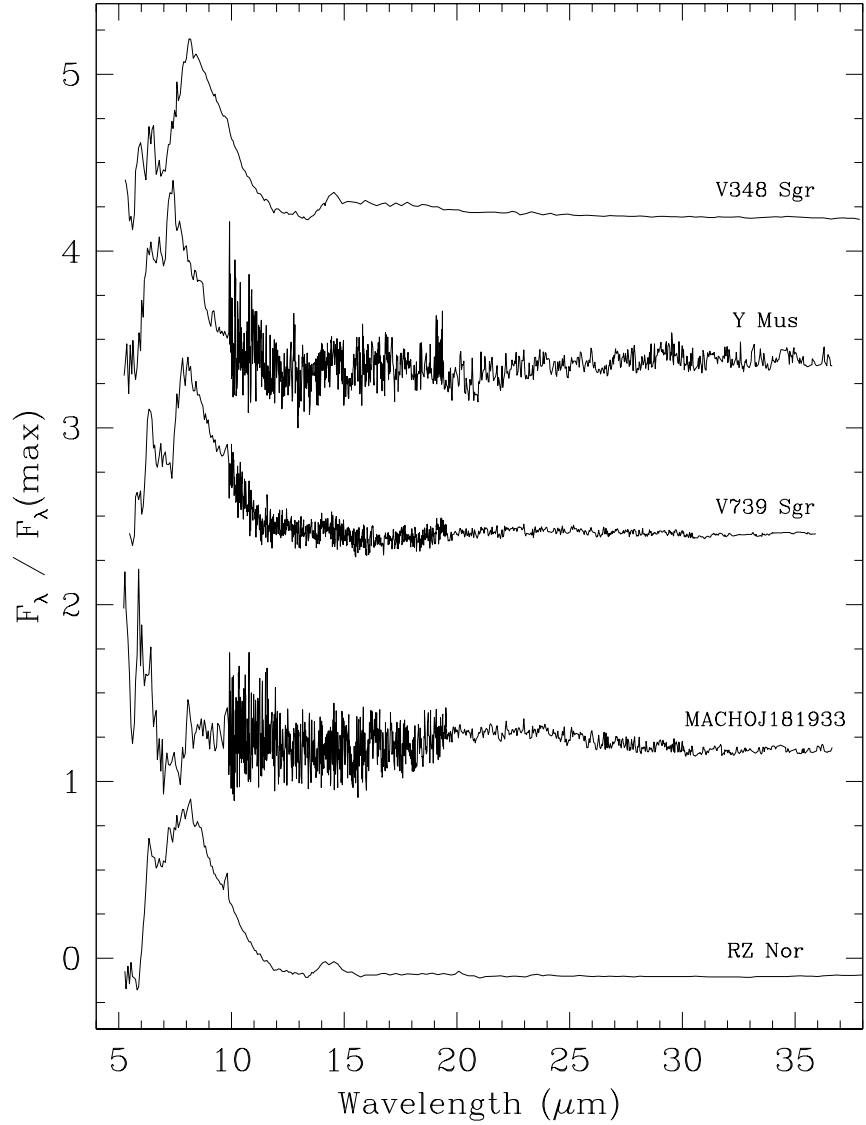}
\caption{{\it Spitzer/IRS} residual
spectra over the full wavelength range
$\sim$5$-$37 $\mu$m for high E(B-V) ($>$0.30) RCB stars V348 Sgr, Y Mus,
V739 Sgr, MACHOJ181933, and RZ Nor, which are ordered from top to
bottom by increasing E(B-V) from 0.45 for V348 Sgr to 0.50 for RZ Nor.
Note that the spectra are normalized and displaced for clarity.
\label{fig5}}
\end{figure}

\clearpage

\begin{figure}
%Figure 6
\includegraphics[angle=0,scale=.60]{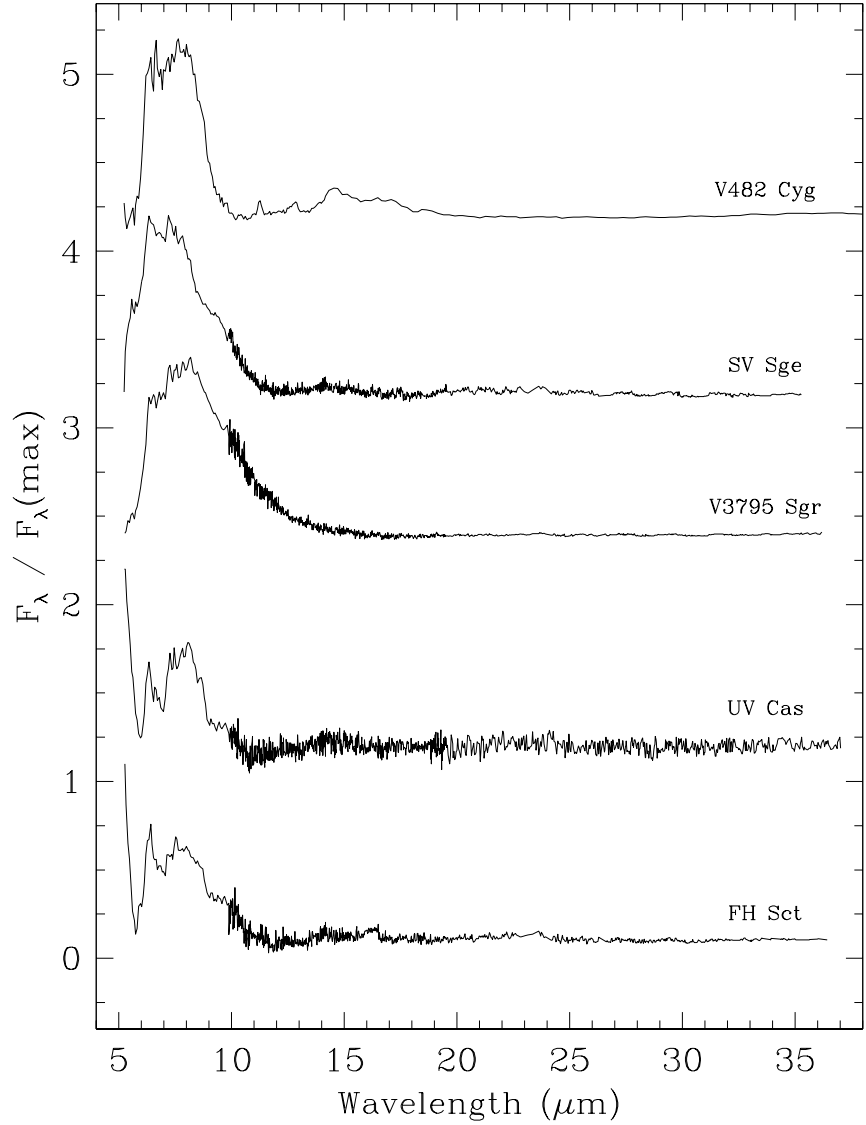}
\caption{{\it Spitzer/IRS} residual
spectra over the full wavelength range
$\sim$5$-$37 $\mu$m for high E(B-V) ($>$0.30) RCB stars V482 Cyg, SV Sge,
V3795 Sgr, UV Cas, and FH Sct, which are ordered by increasing E(B-V) from 
0.50 for V482 Cyg to 1.00 for FH Sct.
Note that the spectra are normalized and displaced for clarity.
\label{fig6}}
\end{figure}

\clearpage

\begin{figure}
%Figure 7
\includegraphics[angle=0,scale=.60]{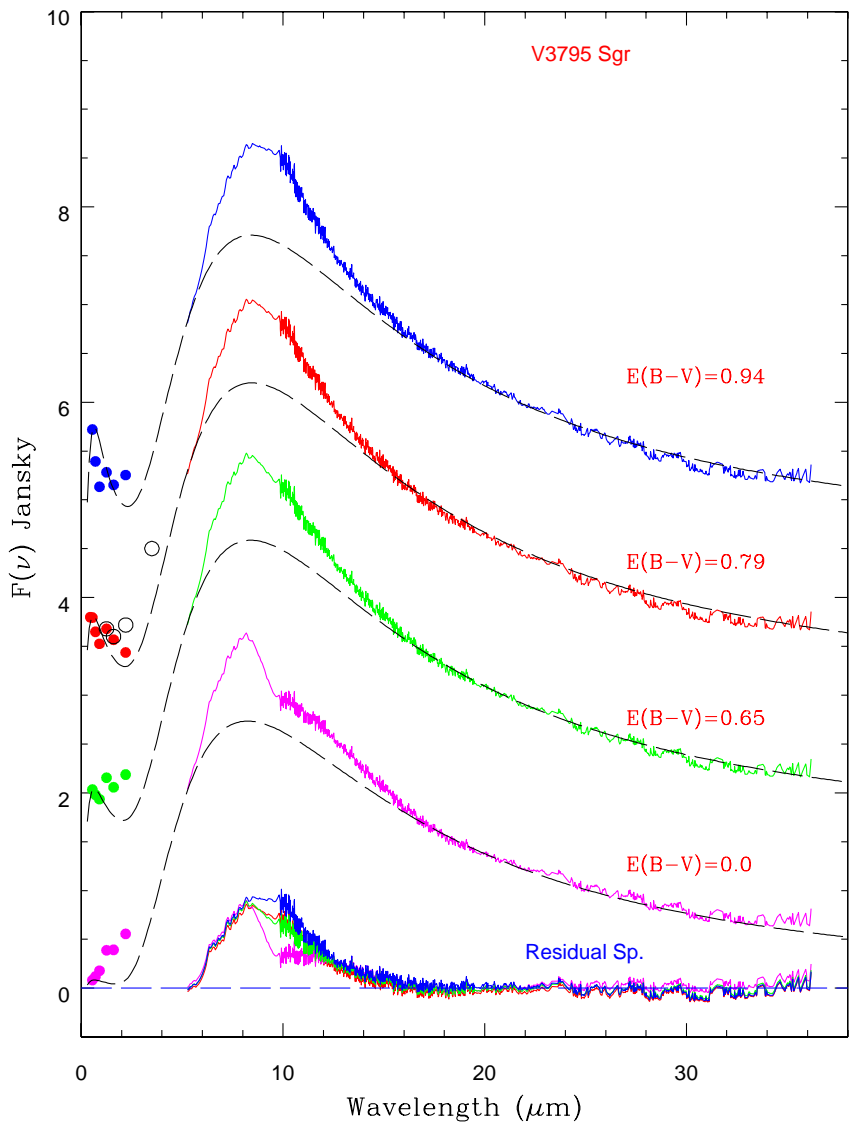}
\caption{SEDs of V3795 Sgr after ISM correction for various E(B-V) values and
fitted with two blackbody temperatures (black dashed lines) representing the
central star (8000 K) and circumstellar dust ($\sim$600-610 K). The dots are the
VRIJHK band flux densities. The SEDs have been shifted by a constant for
clarity. At the bottom of the figure,  the resultant  residual
spectra are  superposed on each other to illustrate the E(B-V) dependence. The
optical and near-IR colors are best matched with the adopted E(B-V) of 0.79.
\label{fig7}}
\end{figure}

\clearpage

\begin{figure}
%Figure 8
\includegraphics[angle=0,scale=.60]{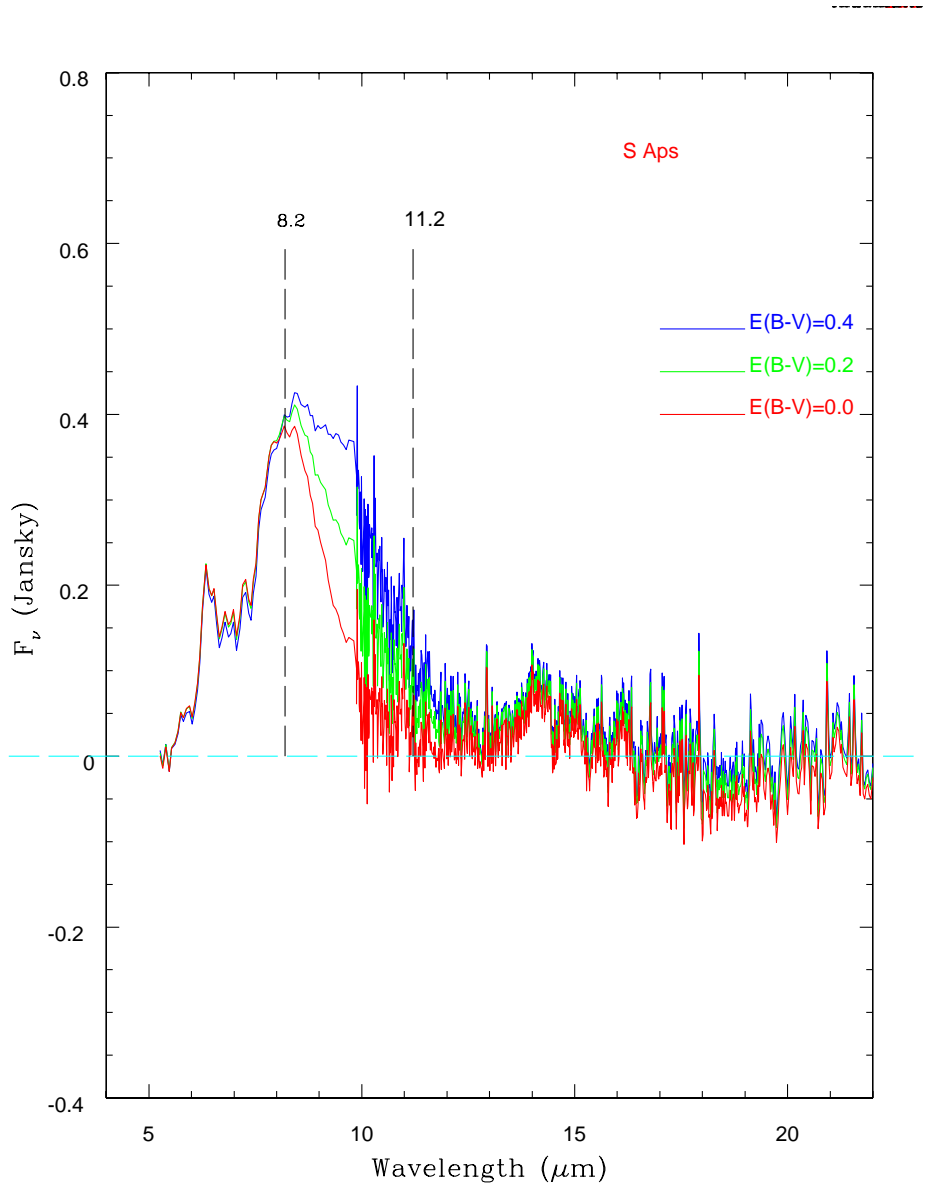}
\caption{S Aps resultant residual
spectra for E(B-V) values of 0.0 (in red), 0.2 (in green), and 0.4 (in
blue). Note that the vertical lines mark the spectral region most affected by
the interstellar reddening correction, specifically the correction for the
interstellar silicate feature (see text).  \label{fig8}}
\end{figure}

\clearpage

\begin{figure}
%Figure 9
\includegraphics[angle=0,scale=.60]{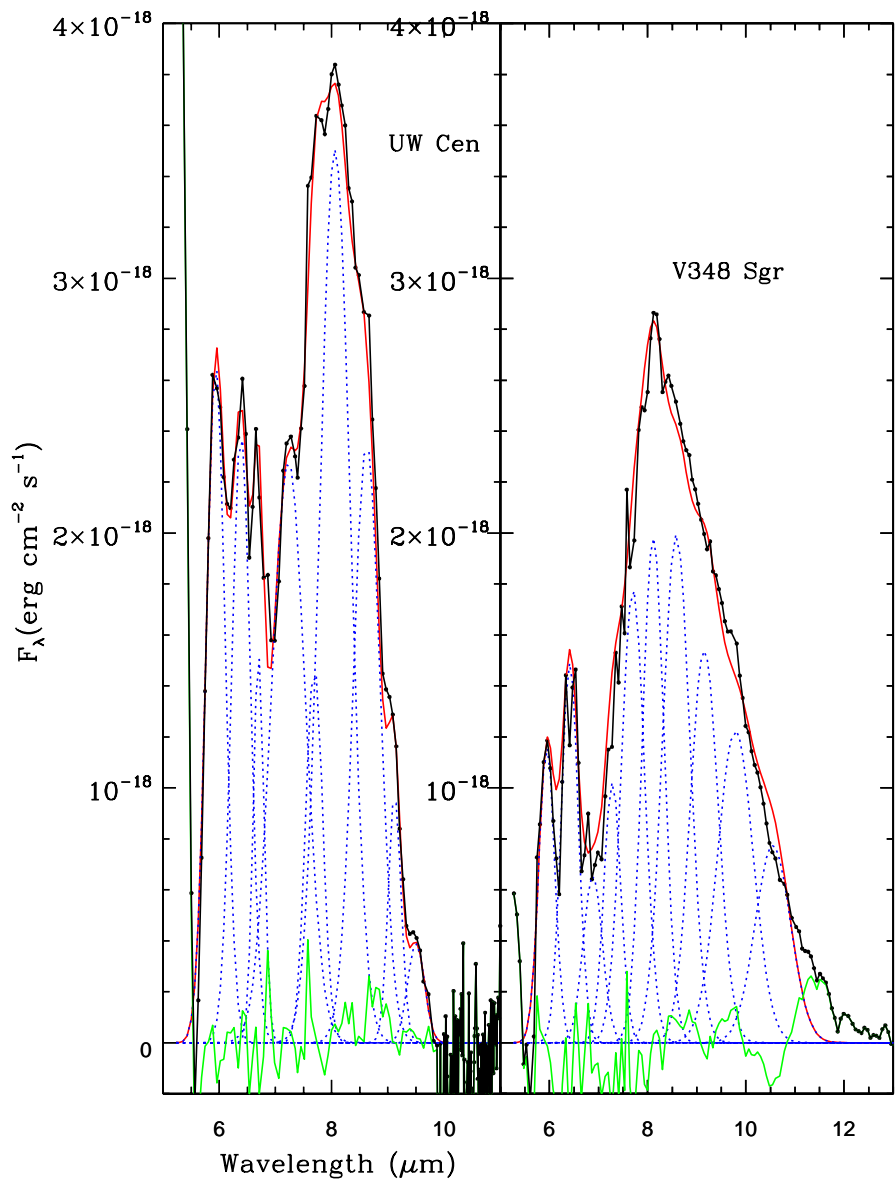}
\caption{Illustrative examples of the multi-Gaussian fits (in red) performed on
the residual {\it Spitzer} spectra (in black) of the warm RCB UW Cen (left
panel) and the hot RCB V348 Sgr (right panel).  Individual gaussian fits (in
blue) around the wavelengths of $\sim$5.9, 6.3, 6.9, 7.3, 7.8, 8.1, 8.6, and
around 9 $\mu$m are shown. The difference between observations and the fit is
displayed at the bottom (in green). \label{fig9}}
\end{figure}

\clearpage

\begin{figure}
%Figure 10
\includegraphics[angle=0,scale=.60]{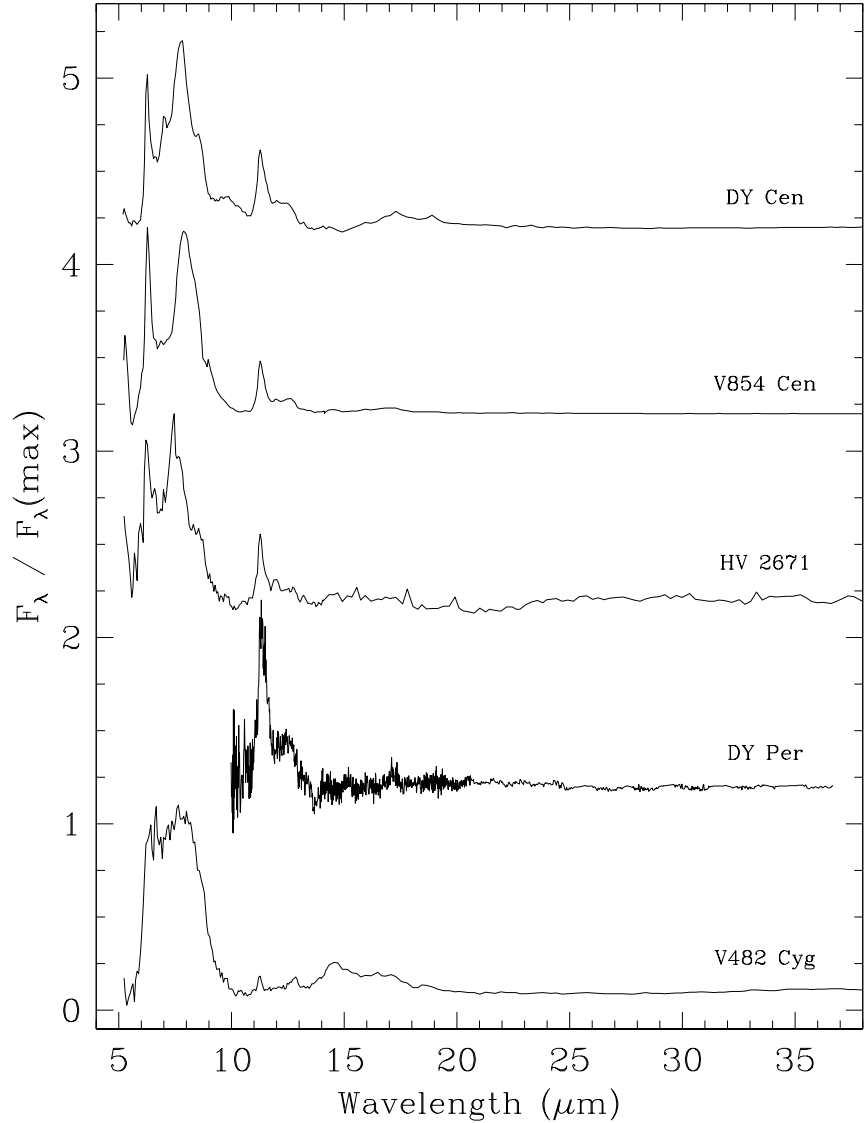}
\caption{{\it Spitzer/IRS} residual spectra over the full wavelength
range $\sim$5$-$37 $\mu$m for the three least H-deficient RCB stars DY Cen, V
854 Cen, and HV 2671 and for DY Per which also shows the 11.3 and 12.7 $\mu$m
PAH-like bands. V482 Cyg's spectrum is shown at the bottom of the Figure. V482
Cyg is an odd case, displaying the typical 6-10$\mu$m dust complex together with
weak 11.3 and 12.7 $\mu$m features (see Section 5). Note that the spectra are
normalized and displaced for clarity. \label{fig10}}
\end{figure}

\clearpage

\begin{figure}
%Figure 11
\includegraphics[angle=0,scale=.60]{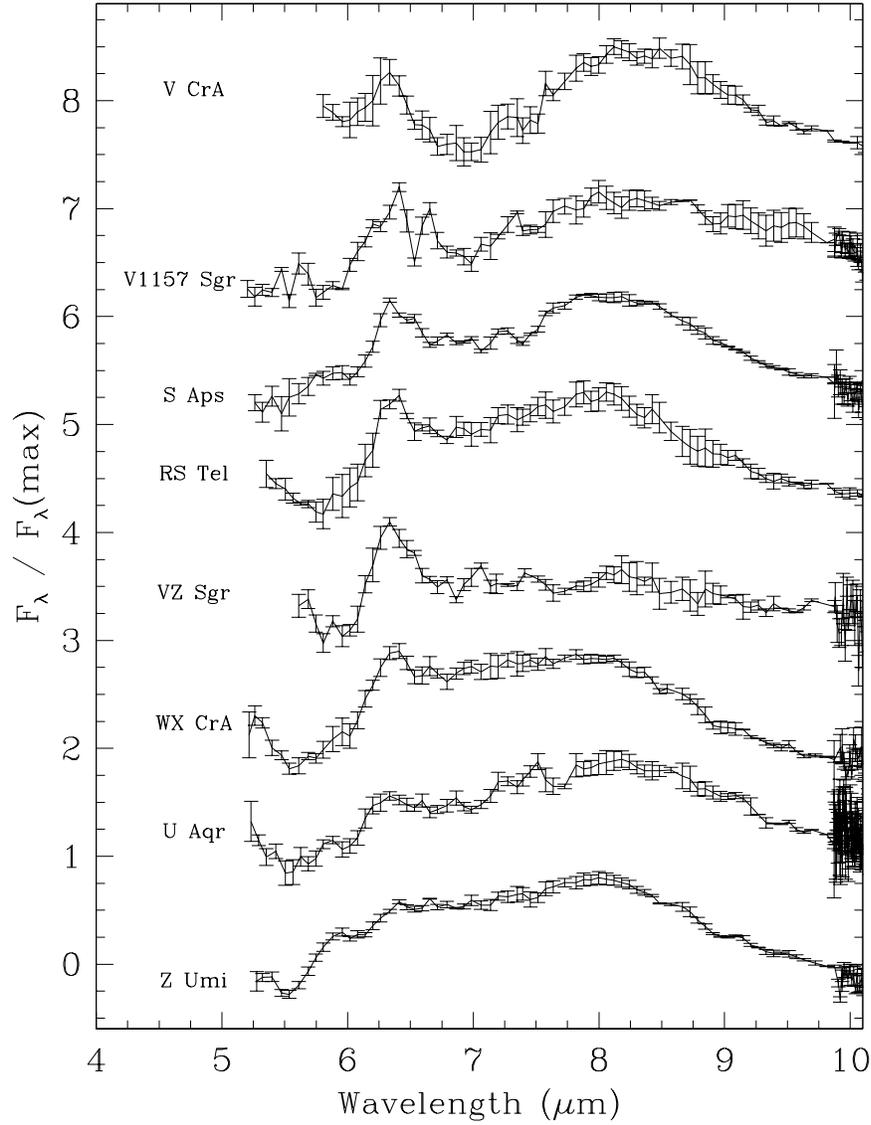}
\caption{{\it Spitzer/IRS} residual
 spectra in the $\sim$5$-$10 $\mu$m
wavelength range for the eight low E(B-V) ($\leq$0.30) RCB stars; from top to 
bottom: V CrA, V1157 Sgr, S Aps, RS Tel, VZ Sgr, WX CrA, U Aqr, and Z UMi. 
Flux error bars are superimposed on the spectra, which are normalized
and displaced for clarity. \label{fig11}}
\end{figure}

\clearpage

\begin{figure}
%Figure 12
\includegraphics[angle=0,scale=.60]{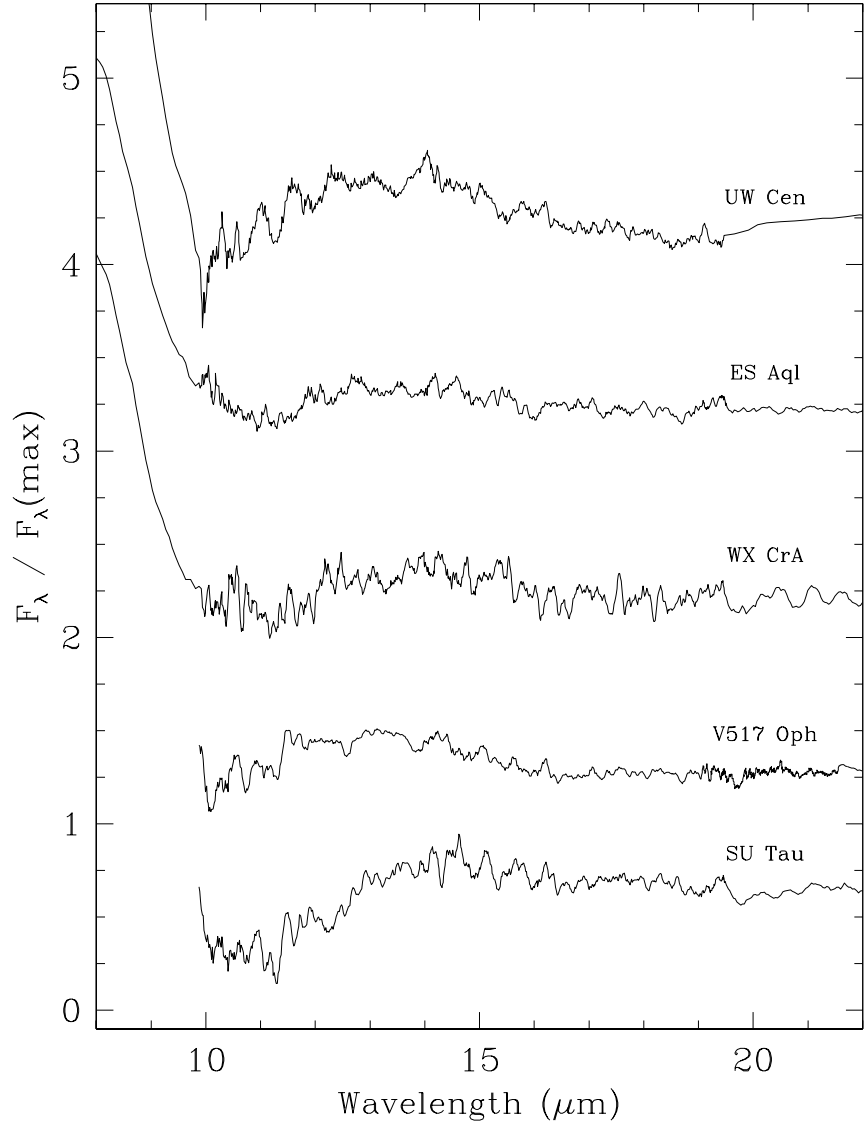}
\caption{{\it Spitzer/IRS} residual spectra in the $\sim$8$-$22
$\mu$m wavelength range for RCBs UW Cen, ES Aql, WX CrA, V517 Oph, and SU Tau.
The residual spectra have been smoothed with boxcar 7 to highlight the very
weak and broad $\sim$11$-$15 $\mu$m emission feature. Note also that the spectra
has been scaled and displaced for clarity. \label{fig12}}
\end{figure}

\clearpage

\begin{figure}
%Figure 13
\includegraphics[angle=0,scale=.60]{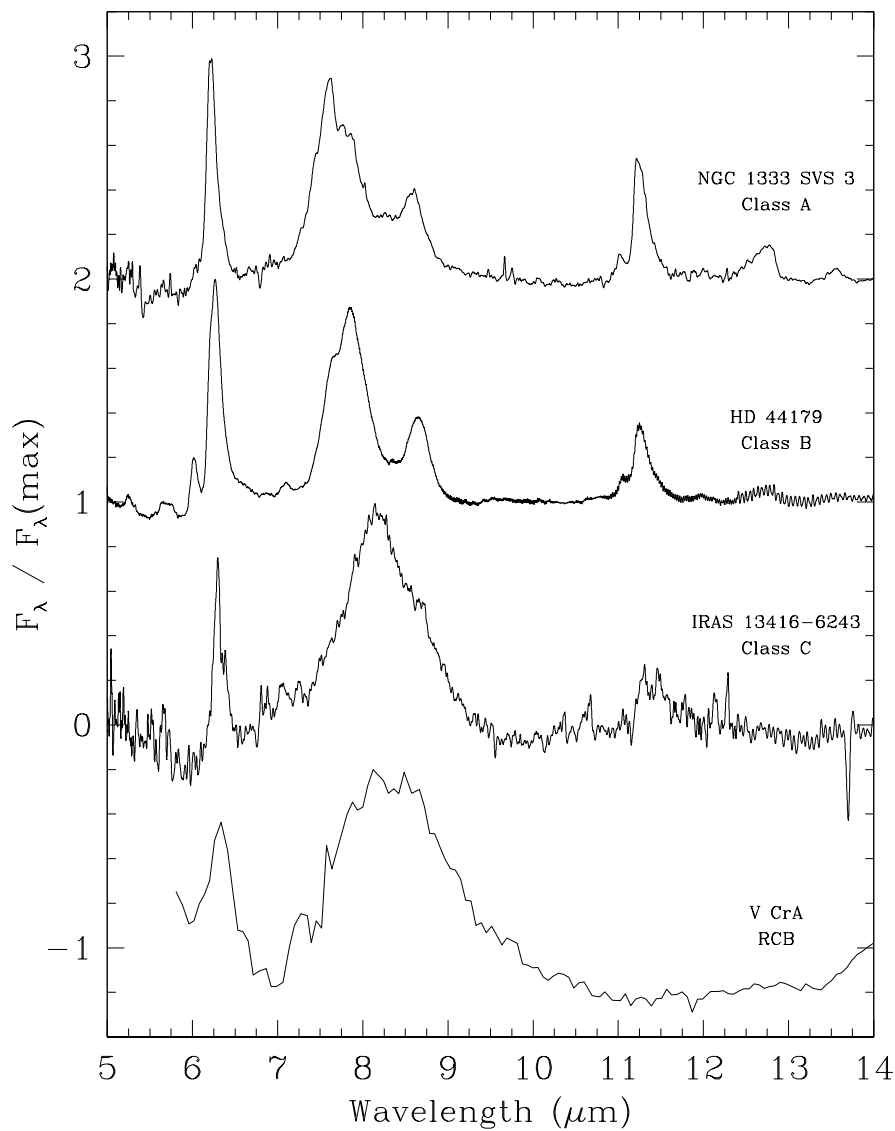}
\caption{Residual ISO spectra representative of Peeters et al.(2002) classes A,
B, and C in comparison with the Spitzer residual spectrum of the RCB V CrA. The
original ISO spectra were taken from Sloan et al. (2003). We subtracted the dust
continuum by fitting 10 degree polynomials in a way similar to Sloan et al.
(2007) (i.e., using similar flux points as dust continuum). Note also that the
residual ISO spectra have been smoothed (with a 15-point box car filter) to be
compared with the VCrA's Spitzer spectrum. \label{fig13}}
\end{figure}

\clearpage

\begin{figure}
%Figure 14
\includegraphics[angle=0,scale=.60]{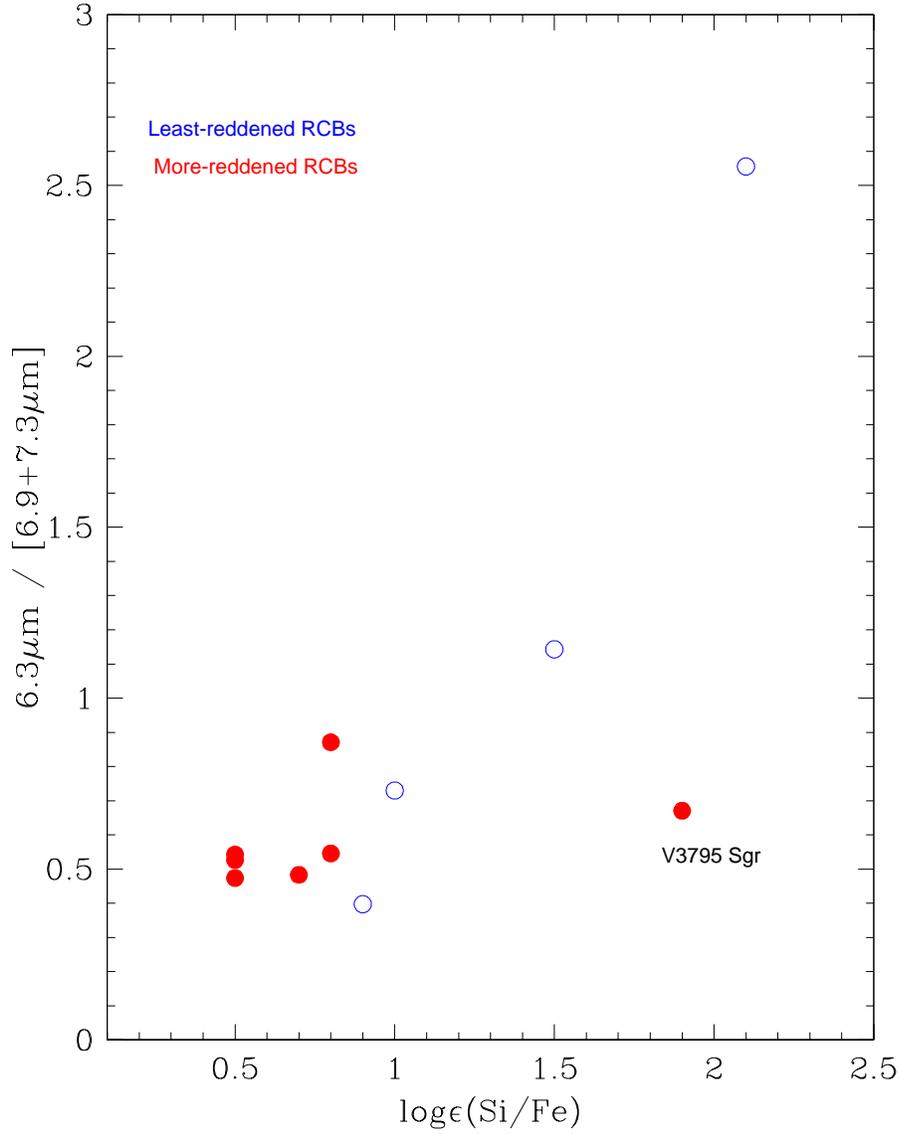}
\caption{Flux ratio (6.3$\mu$m/6.9-7.3$\mu$m) versus the silicon abundance
(log$\epsilon$(Si/Fe) taken from Asplund et al. (2000).  The least- and
more-reddened RCBs are marked with blue and red symbols, respectively. The
`minority' RCB V3795 Sgr with a rather exceptional infrared spectrum is 
labelled. \label{fig14}}
\end{figure}

\clearpage

\begin{figure}
%Figure 15
\includegraphics[angle=0,scale=.60]{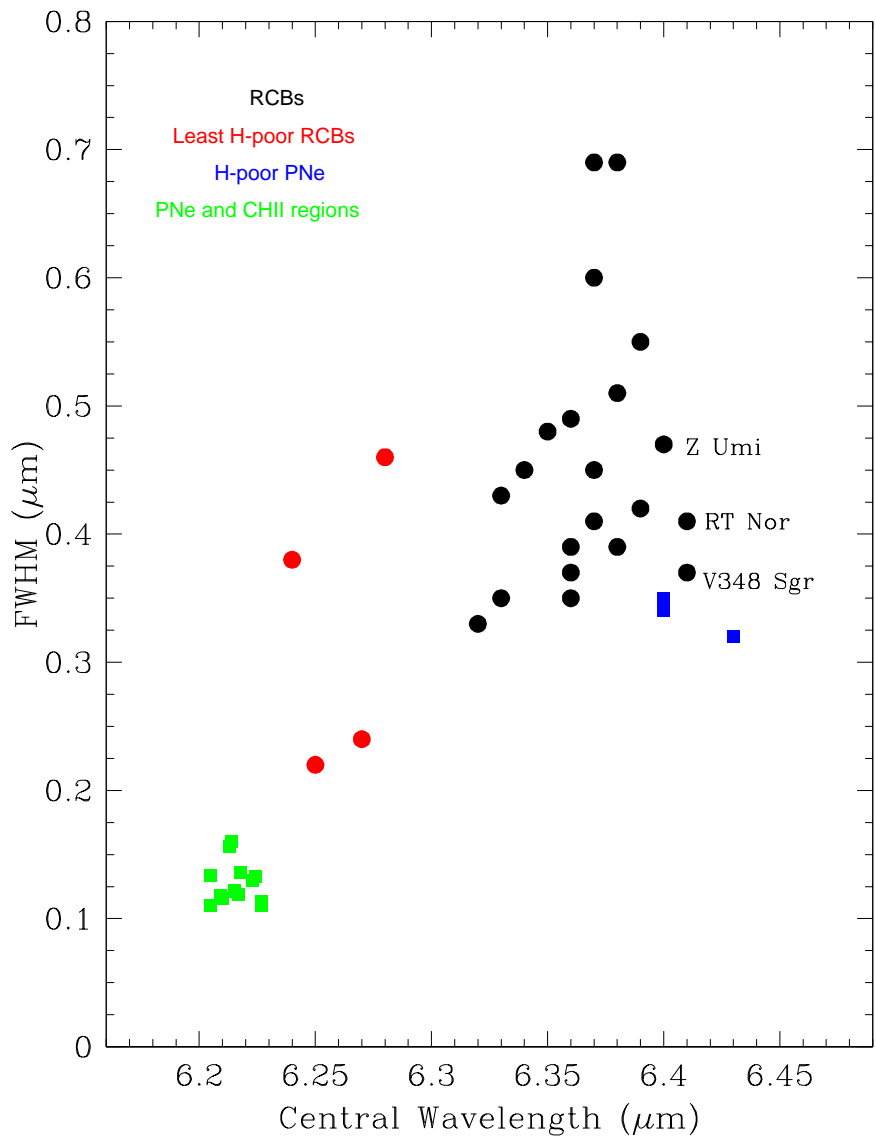}
\caption{Full width at half maximum (FWHM) versus the central wavelength for the
6.3 $\mu$m feature observed in RCB stars (black circles). This feature is
blue-shifted to $\sim$6.26 $\mu$m in the least H-deficient RCBs (DY Cen, V854
Cen, HV 2671, and V482 Cyg, red circles) but red-shifted to $\sim$6.41 $\mu$m in
H-poor PNe (blue squares). The 6.3 $\mu$m feature is blue-shifted to $\sim$6.22
$\mu$m in normal H-rich PNe and compact H II (CHII) regions (green squares;
Peeters et al. 2002). Note that a few RCBs showing a red-shifted feature similar
to the H-poor PNe are labelled. 
\label{fig15}}
\end{figure}

\clearpage

\begin{figure}
%Figure 16
\includegraphics[angle=0,scale=.60]{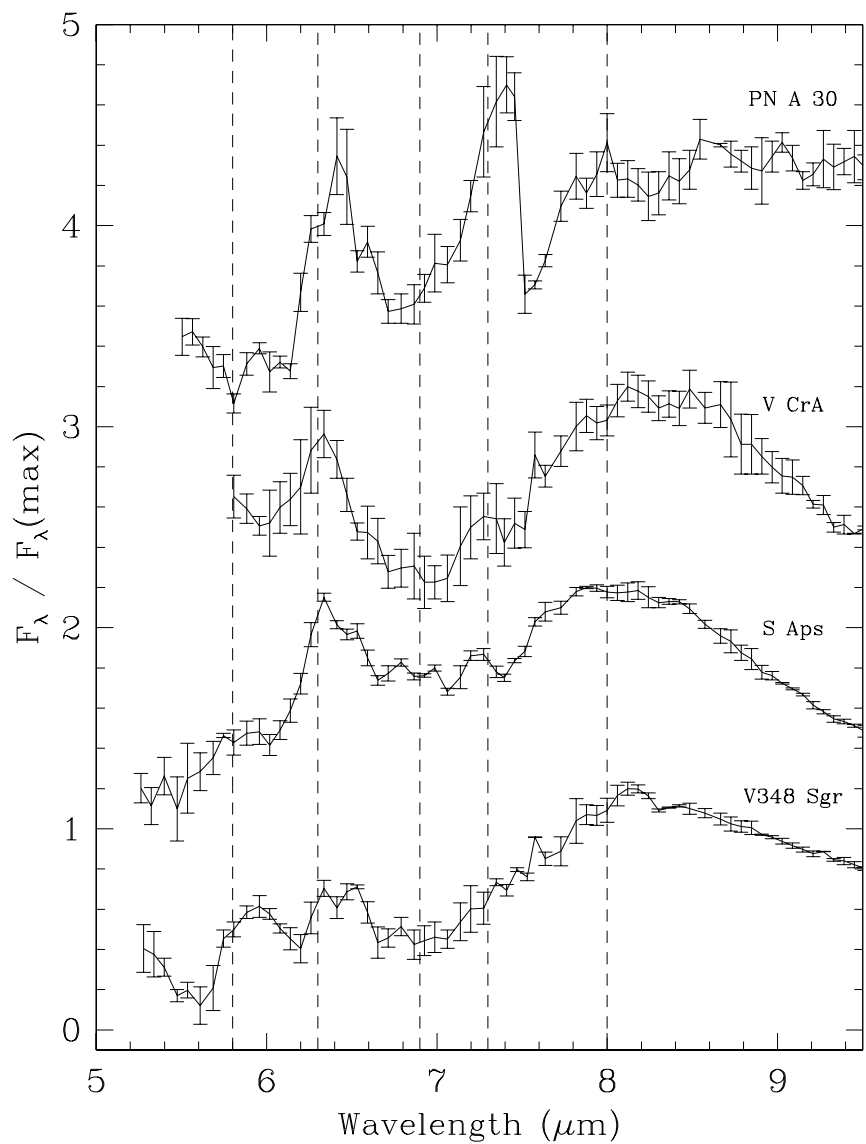}
\caption{{\it Spitzer/IRS} residual
 spectra in the wavelength range $\sim$5$-$10
$\mu$m for the H-poor PN A 30 compared with the RCB stars V CrA  (a `minority'
RCB), S Aps (a cool RCB), and V348 Sgr (a hot RCB). Wavelengths of
the infrared bands of amorphous carbon grains at 5.8, 6.3, 6.9, 7.3 and 8.0
 (Colangeli et al. 1995) are marked with black dashed vertical
lines. Note that the spectra are normalized and displaced for clarity.
\label{fig16}}
\end{figure}

\end{document}